\documentclass[]{interact}

\usepackage{epstopdf}% To incorporate .eps illustrations using PDFLaTeX, etc.
\usepackage[caption=false]{subfig}% Support for small, `sub' figures and tables
\usepackage{amssymb}
\usepackage{lineno,hyperref}
\usepackage[table]{xcolor}
\usepackage{multirow}
\usepackage{subfig}

\usepackage[numbers,sort&compress]{natbib}% Citation support using natbib.sty
\bibpunct[, ]{[}{]}{,}{n}{,}{,}% Citation support using natbib.sty
\renewcommand\bibfont{\fontsize{10}{12}\selectfont}% Bibliography support using natbib.sty
\makeatletter% @ becomes a letter
\def\NAT@def@citea{\def\@citea{\NAT@separator}}% Suppress spaces between citations using natbib.sty
\makeatother% @ becomes a symbol again

\theoremstyle{plain}% Theorem-like structures provided by amsthm.sty

\theoremstyle{definition}

\theoremstyle{remark}

\begin{document}

%\articletype{ARTICLE TEMPLATE}% Specify the article type or omit as appropriate

\title{The epidemiology of lateral movement: exposures and countermeasures with network contagion models}

\author{
\name{Brian A. Powell\thanks{Email: brian.a.powell@jhuapl.edu}}
\affil{Johns Hopkins University Applied Physics Laboratory\\11100 Johns Hopkins Rd., Laurel, MD 20723}
}

\maketitle

\begin{abstract}
An approach is developed for analyzing computer networks to identify systems and accounts that are at particular risk of compromise by an adversary seeking to move laterally through the network via authentication. The dynamics of the adversary are modeled as a contagion spreading across systems linked via authentication relationships derived from Administrator account access and active session data. The adversary is assumed to traverse the network via credential chaining, where the adversary steals credentials from one system, uses them to authenticate to another, and repeats the process. Graph topology measures are used to analyze different contagion models applied to a real Windows network for three primary exposures by identifying: accounts which, either individually or collectively, provide wide and far-reaching access to many systems across the network; accounts with notable privilege escalation liability; and ``gatekeeper'' systems through which the adversary must pass in order to reach critical assets. The approach can be used to test how different mitigations and countermeasures affect these exposures; for example, we find that disabling remote logins by local accounts and implementing protections that prevent the caching of credentials on remote hosts can substantially curtail lateral movement and privilege escalation.
\end{abstract}

\begin{keywords}
network security; lateral movement; credential theft; privilege escalation
\end{keywords}

\section{Introduction}
Cyber attackers are almost always hunting for something---credit card numbers, password files, or the highly-guarded recipe for the secret sauce. This data is ideally tucked away somewhere, out of reach of all but the most trusted individuals.  When attackers penetrate a network, the compromised ingress system usually does not hold the secret data, and the compromised account might not have access to the system that does.  And so begins the delicate process of {\it lateral movement}, as the adversary steps through the network, implanting itself, hunting for sensitive data, seeking to escalate privileges along the way in order to access ever more secure systems.  The key to lateral movement is to keep exceptionally quiet so that the adversary has time to find what it's looking for; stealth is achieved by getting lost in the noise, by blending into the normal rhythm of the network.  This means that unauthorized activities should be kept to a minimum: the adversary must ``live off the land'', using native utilities already present on compromised hosts to gain persistence, perform reconnaissance, and make remote connections.  Movement from system to system should be {\it authenticated}, so that it is appears authorized and otherwise unremarkable.  To fuel this authenticated lateral movement, the adversary must harvest credentials from compromised hosts as quietly as possible, stringing together compromised accounts and escalating privilege until the target is reached.  This is actually all a lot easier than it might sound.

Once the adversary moves off its ingress point and into bustle of the network, they tend to dwell undetected for a long time---a 2017 study \cite{Ponemon} reports an average of 191 days.  If the hope is to catch them before they disappear, recent estimates \cite{Crowdstrike} place the average time between ingress and first lateral move at two hours---not a lot of time to detect and respond to all but the noisiest of intruders.  Much effort has been put towards curtailing lateral movement by locking down native tools and securing credentials, but such safeguards are not a cure-all: organizations also need to ensure that users don't have unneeded access or excessive privileges, and that Administrators observe best practices like least privilege and avoid crossing security tiers.  Defenders must adopt a proactive strategy of hunting for unusual logins, suspicious host processes, and evidence of credential theft.  The trouble is that enterprise networks are large,  with often complex access control policies that are difficult to audit and analyze, and defenders generally don't know where to look to catch lateral movement.

In this paper, we develop an analytical framework that can be applied to any computer network to gain insight into these challenges. Using Administrator account access data, we map out the basic authentication relationships of all systems across the network: essentially, which system's Local Administrator accounts have privileges on which other systems, forming a directed graph called the {\it authentication network}, or authnet.  A Local Administrator is any account belonging to a system with sufficient privileges to perform restricted administrative tasks.  We employ a simple contagion model to emulate an adversary seeking to expand access via the basic technique of {\it credential chaining}, in which the adversary steals credentials from one system, uses them to authenticate to another, steals credentials there, and continues in this manner. We quantitatively asses this model using a select set of graph centrality measures to identify systems that are especially good {\it spreaders}---good facilitators of lateral movement.   We find that often such systems are unremarkable individually, but are connected together in ways that confer wide and far-reaching access across the network.  Results like these can yield insights into access control configuration and help identify accounts with too much access, either individually or collectively.

We next examine the role of {\it active sessions} on the network. An active session is the temporary access of an information system by a user following a successful logon event.  When users login to a computer their credentials are generally available on that system, either in running memory or cached in the security database, until logout or system reboot.  In this work we refer to the general availability of credentials, of any type, in running memory and cache as {\it credential residue}, to distinguish them from credentials residing indefinitely in local security databases.  Credential residue is the primary source of credentials of high-privileged network accounts that are not local to any system (and so don't have credentials stored in any local security database) contributing to privilege escalation exposure. Using login data from authentication logs, we map these authentication relationships on top of the Local Administrator authnet to obtain a more comprehensive picture of all possible authenticated routes through the network.  We analyze this authnet to identify and characterize cases of cross-tier access by Administrator accounts in different security tiers: for example, an Administrator in the highest security tier accesses an asset that is also accessible by a lesser-tier Administrator that has a concurrent active session on the network.  An adversary can escalate privileges by compromising the lesser-tier Administrator's session, called an {\it escalator}, and authenticating to the system hosting the higher-tier Administrator's active session. These findings are useful for monitoring cross-tier activity, and can help craft policy that limits privilege escalation liability.

Lastly, we explore how to defend certain high value systems from unauthorized authenticated access.  Given a set of such systems, the authnet is used to enumerate all systems that can reach them via chained authenticated access: these access paths can be generated in near-real time so that defenders can shift and adapt to the changing network. We apply a contagion model to this network that incorporates the possibility of recovery (detection) in order to identify the major conduits, called {\it gatekeepers}, through which the adversary must pass en route to the targets.  By focusing on this small set of key systems for more careful monitoring, cyber defenders can anticipate the adversary by widening the defensive boundary a step or two out from critical assets.  

%Though we employ contagion models to enumerate the possible pathways the adversary might take through the network, and so identify high-risk systems, we also develop a parallel approach that achieves the same objective more easily by applying graph analytics to the authnet itself.  Depending on the application (finding spreaders, escalators, or conductors), an appropriate set of graph centrality measures is applied to the authnet such that the centrality of each vertex approximates its exposure according to the epidemeological model.  Then, a clustering-based outlier technique is applied to the graph centrality space to identify exceptional systems. Though lacking in conceptual richness, this latter approach is easier to implement and so is the recommended application.  

This analysis affords engineers and administrators the ability to test different network-wide or account-specific countermeasures against the threat of lateral movement.  For example, once the authentication network is established, it is possible to explore what happens to spreadability when certain types of inter-system remote access are restricted, or what effect restrictions on in-memory credential storage have on privilege escalation.  We explore a few of these mitigations in this paper to demonstrate the methodology.   

In what follows, we collect data from and apply these methods to a real Windows Domain.  Hence, the derived authentication relationships embodied in the authnet are based on the details of specific Windows authentication protocols, logon processes, and so on; the specifics of the Windows operating system and authentication processes that are prerequisites for these applications are provided in Appendix A.  Once the authnet is built, however, the analytic methods are general and can be applied to any computer network.   

Throughout this paper numerical values for things like network size, number of active Administrator sessions, or the percentage of escalators eliminated by a certain countermeasure will be reported; since these numbers are specific to one particular network, they are of no interest to a general audience.  The emphasis is therefore not on the numbers or the characteristics of this particular network, but on the methods of analysis that render them. By working through an analysis of a real network, we hope to demonstrate how the methods described here can be used by security analysts to explore, understand, and better defend their networks. 
\section{Cyber Attack as an Epidemic}
The formal basis for this analysis is the study of epidemic spread as percolation in networks \cite{Newman,Newman2,Meyers,Sander}.  In contrast to traditional epidemiological dynamics in which any infected individual in a population can make contact with, and potentially infect, any other individual in the population, the spread of epidemics on networks is constrained by the network structure---infected vertices can only spread contagion to those vertices with which they share an edge, and, on directed networks, only in the direction of the edge. This formalism has found applications in biology, social networks, and, increasingly, cybersecurity. 

Epidemic modeling classifies hosts as susceptible, infected, recovered (with or without immunity) or dead.  Models are named for the states involved, {\it e.g.} SI models include only susceptible and infected hosts---there is no death or recovery.  In SIR models, infected hosts can recover with immunity, so that they cannot be reinfected; in contrast, SIRS models include the possibility of reinfection by assuming that immunity is only temporary.  

%The contagion that evolves in these models is an unthinking, non-deliberate agent, quite unlike the real cyber adversary that anticipates, reacts, and makes use of strategy.  For this reason, the application of epidemiology to cybersecurity has generally been limited to modeling the spread of equally unthinking entities, like computer viruses and worms \cite{Kephart,Wang,Zou}.  However, in studying the key systems that generally facilitate lateral movement, none of these characteristics are important: absent a specific threat strategy or target set, we let the topology of the authentication network determine we care only to find the most central systems, which due to the structure of the authentication network, will be visited most often on average by any adversary following any strategy (except for a strategy that explicitly calls for avoiding all such systems, a rather unlikely scenario).   The unrelenting spread of a mindless contagion across the authentication network thus allows us to model a generic adversarial lateral movement. Later on, when we analyze gatekeeper systems, we allow the adversary to adopt a targeted strategy by allowing the contagion to spread only along shortest paths; otherwise, it too is unguided so that we can fully explore the disposition of the network to a generic, unspecified threat. 

We model adversarial lateral movement as a contagion spreading on a directed network, with infection of a vertex corresponding to adversarial access to the corresponding system.  Infection can only spread between vertices in the direction of the edge, in accordance with the authentication relationship between the connected systems.  We assume that infected systems always infect susceptible systems with certainty and that the contagion spreads serially, one host at a time, much like the true cyber adversary. A single {\it epidemic scenario} corresponds to the sequence of infections that follow from a single initially-infected host.   We assume that the adversary retains access to all systems accessed during the scenario, and, hence, that systems remain infected for the duration of the epidemic scenario.  Each scenario runs until no further hosts can be infected (based on network topology), or until it is eradicated, which corresponds to the simultaneous recovery of all hosts; in our cybersecurity analogy, once the adversary is detected on a system, the attack is considered over. 

For each application, we consider an ensemble of epidemic scenarios.  To study credential exposure across the entirety of the network, we run scenarios until each vertex has been infected at least once.  The ensemble of scenarios is represented as an {\it infection network}, which includes all systems infected in the ensemble as vertices and all transmission paths as edges. The in-degree of a vertex gives the number of times it was infected by a distinct host, the out-degree gives the number of distinct hosts infected, and the weight of an edges gives the number of times the contagion spread between the associated vertices.  We apply a variety of graph centrality measures and other metrics to the infection network in order to identify important systems. 

Contagion has been used in a number of other cybersecurity studies.  Epidemic models have been useful in modeling the spread of unthinking computer viruses and worms \cite{Kephart,Wang,Zou}. In \cite{Acemoglu}, the effect of system security investments on infection likelihoods is studied; the adversary is modeled as a contagion spreading probabilistically across the network according to the encountered security investment of each system.  The work of \cite{Goyal} explores a slightly more intelligent adversary, which deploys attack resources on systems in a bid to capture them according to a probabilistic contest.  The objective is to identify the optimal network structure that reduces the potential spread of the adversary, which is essentially modeled as a contagion that simultaneously spreads to multiple uninfected systems as resources allow.  Contagion models have also been used to study network resiliency \cite{Shao,Blume}, where an attack causing system failure cascades through the network spreading indiscriminately from infected to susceptible systems.

There are also some interesting results from graph percolation theory that are relevant to the spread of infections on networks.  In particular, in \cite{Piraveenan} the notion of {\it percolation centrality} is introduced in order to ascertain vertex centrality in the presence of a percolation process.  Evaluation of this metric requires knowledge of the percolation state of vertices in the network, which, in cybersecurity applications, corresponds to a knowledge of which systems are compromised.  This metric is especially useful for determining at-risk vertices in the presence of a particular, ongoing infection.  In contrast, here we consider all possible, hypothetical epidemics on the authnet in order to identify those systems that are generically vulnerable to compromise.  
\section{Finding Spreaders}
The first exposure we investigate is {\it spreadability}, which is the tendency of an account to provide significant access to other systems, either immediately or collectively through chained access.  The basis for this analysis are the authentication relationships of Local Administrator accounts arranged as a directed graph called an {\it authentication network}: we now describe the structure of this graph and the data sources needed to create it.  
\subsection{Local Administrator Authentication Network}
The authentication network, or authnet, is a directed graph: vertices are information systems and edges indicate authentication relationships between them.  Specifically, an edge $E = (u,v)$ from vertex $u$ to vertex $v$ indicates that a credential available on the system represented by vertex $u$ can authenticate with Administrative privileges to the system represented by vertex $v$.\footnote{Strictly speaking, $u$ and $v$ are generic labels of the vertices {\it representing} systems, but, for brevity, in what follows we will consider systems and the vertices representing them as one and the same.}  Given two edges, $E = (u,v)$ and $E' = (v,w)$, the directed path $E \cup E' = (u,v,w)$ can be traversed via {\it credential chaining}, that is, by using the credential available on system $u$ to authenticate to system $v$, followed by the use of the credential available on $v$ to access system $w$. Such paths can be arbitrarily long.  The local account on system $u$ is said to be a {\it derivative Local Administrator} of system $w$.  

%As discussed, credentials can exist on a system in different forms; for example, local account credentials are stored in local databases,  while accounts with active logon processes running on the system, like a remote desktop session, might have relevant credentials available in running memory for at least the duration of the process\footnote{The availability of credentials in these forms is essential to the authentication processes of most operating systems, and though credentials might be found elsewhere, for example in encrypted text documents, we don't account for these non-essential sources.}.  

\begin{figure*}[htp]
\centering
\includegraphics[width=0.85\textwidth,clip]{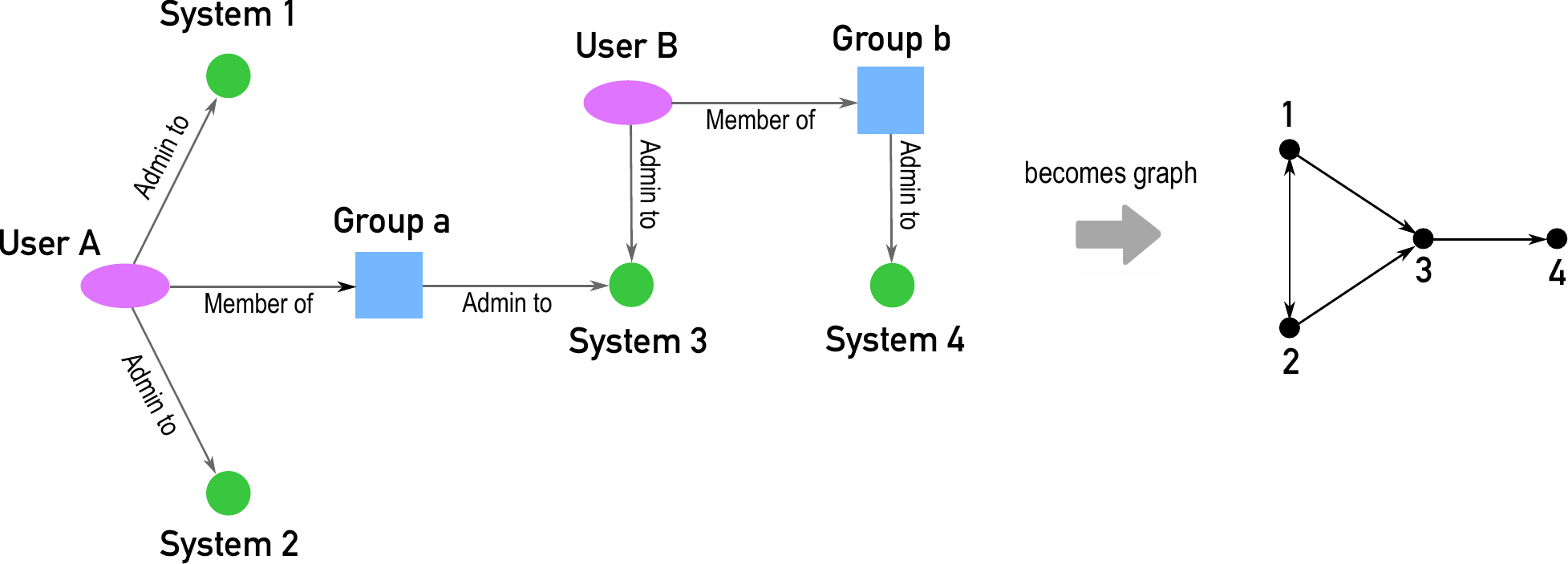}
\caption{Illustration of how account Administrative and group membership relationships link systems, and the directed system graph resulting from this relationship data.}
\label{1}
\end{figure*}

We are first interested in charting the authentication relationships of Local Administrator accounts, and the resulting network of such relationships is a directed graph called the Local Administrator authnet.  To generate the graph, all local accounts in the network must be enumerated, and, for each account, all systems to which it has Administrative access, either directly or through group membership, are identified (see Figure \ref{1}).  Direct Administrative access is represented by an \texttt{Admin to} edge connecting a user and system, whereas if the access is granted through group membership, the user is connected by a \texttt{Member of} edge to a group, and the group is connected by an \texttt{Admin to} edge to the system.

On a Windows network, these data can be acquired from the Active Directory configuration using \texttt{net} commands or through PowerShell with the Active Directory module.  But, it is more convenient to run a few routines from the PowerSploit \cite{PowerSploit} suite: specifically, first retrieve a list of all systems in the Domain with \texttt{Get-DomainComputer}, and then for each system, list members of the Local Administrator group with \texttt{Get-NetLocalGroupMember}.  There will possibly be both user accounts and Domain groups in a given system's Local Administrator group: for each Domain group,  list members with \texttt{Get-DomainGroupMember}.   All of these steps (and more) are bundled conveniently in the Bloodhound Domain reconnaissance tool \cite{BloodHound}: indeed, the Neo4J graph database produced by Bloodhound is in ready form for the analysis depicted in Figure \ref{1} that underlies the authnet. 

\begin{figure*}[htp]
\centering
\includegraphics[width=0.85\textwidth,clip]{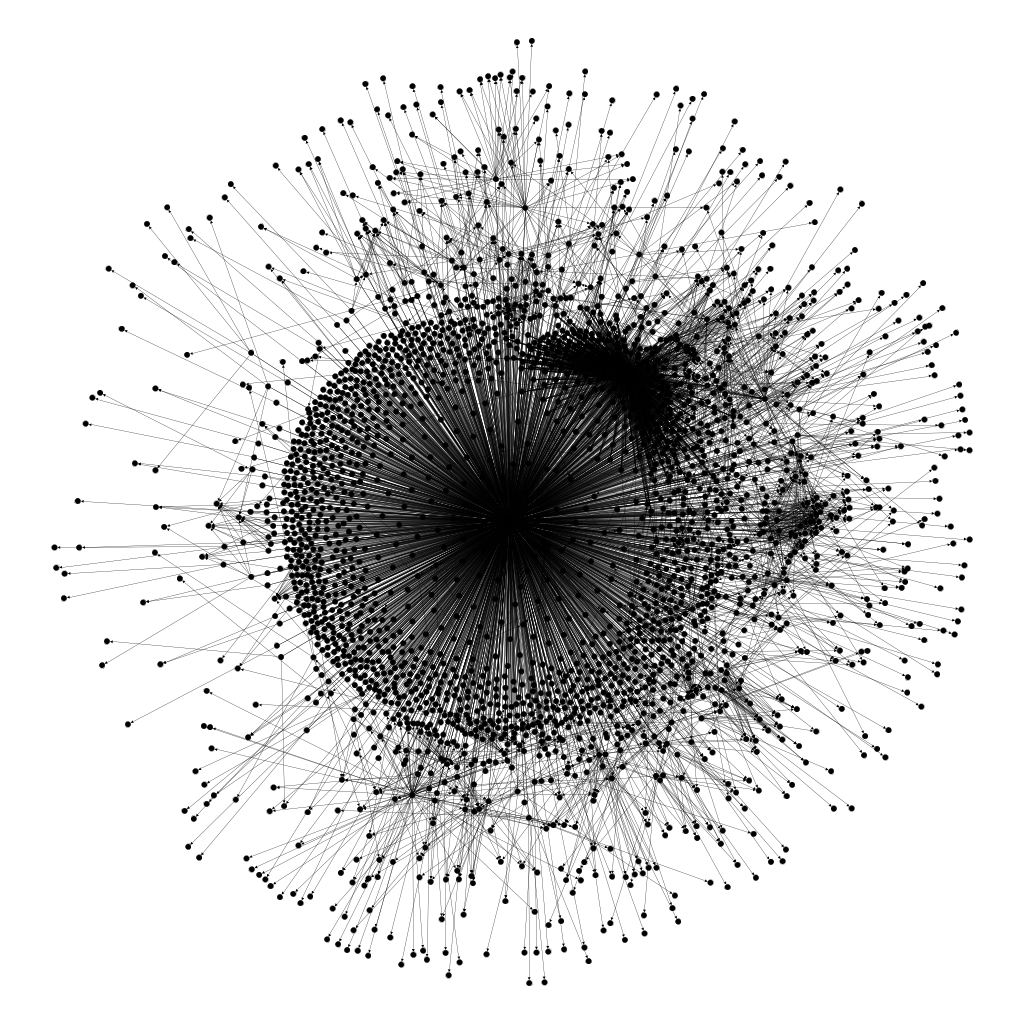}
\caption{Local Administrator authentication network, after clustering similar systems (see text). This and the other graphs in this work were created with the Gephi graph visualization platform \cite{Gephi}.}
\label{2}
\end{figure*}
The Local Administrator authnet of our Windows network includes 3791 vertices, with each vertex representing a unique system in the Active Directory database.  The trouble with this representation is that a single user might have multiple systems associated with their account, {\it e.g.} retired systems still in Active Directory or travel laptops, not all of which are present on the network at the same time.  We therefore cluster those systems deemed to belong to the same user so that they are all represented by a single vertex in the authnet.  Further, some user accounts have access to subnetworks of similar systems, {\it e.g.} nodes in a compute cluster, which from a data perspective are not expected to differ significantly, and offer similar prospects for lateral movement positioning.  We therefore also cluster such similar systems if they have the same parent and child vertices. 

After clustering, the original authnet is reduced to $|V| = 3389$ vertices, 889 of which are singletons with the remainder forming a weakly connected giant component of size $|GCC| = 2499$ (see Table 1).  
The giant component is shown in Figure \ref{2}: it is anchored by two systems, in the center of the graph,  with exceptionally high in-degrees of $k_{\rm in} = 1853$.
Though the giant component holds $70\%$ of the graph's vertices, it is highly fragmented in that it is made up of a patchwork of many small trees, with each tree made up of all reachable vertices from its root (which has $k_{\rm in} = 0$).  The rest of the vertices outside the giant component are singletons---systems with local accounts that cannot authenticate to any other system.  These properties result in a small average eccentricity of less than one hop, and an average number of descendants of just under three.  In short, it is hard to move around appreciably via chained authentication in this network.  Interestingly, though, around $20\%$ of the giant component can be broken off by the removal of just a single articulation point, indirectly indicating the presence of highly connected vertices.  These are some of the vertices we are interested in finding.

\begin{table*}
\footnotesize
\begin{center}
\begin{tabular}{|c|c|c|c|c|c|c|c|c|c|}
\hline
authnet&$|V|$&$|E|$&$\overline{k_{\rm out}}$&Trees&Components&$|RGB|/|GCC|$&$\overline{\epsilon}$&$d$&$\overline{|V_{\rm desc}|}$   \\ \hline
local Admin&3389&8747&2.6&1826&890&0.82&0.59&5&2.9 \\
\hline
%\vtop{\hbox{\strut local account}\hbox{\strut + sessions}}&3716&280617&132&0.007&0.45&133&56&0.97&0.7&7&159&145 \\
restricted&\multirow{2}{*}{3514}&\multirow{2}{*}{6590}&\multirow{2}{*}{1.8}&\multirow{2}{*}{1777}&\multirow{2}{*}{1566}&\multirow{2}{*}{0.97}&\multirow{2}{*}{0.51}&\multirow{2}{*}{2}&\multirow{2}{*}{1.87}\\
local Admin&&&&&&&&&\\
\hline
\end{tabular}
\end{center}
\caption{Local Administrator authnet characteristics.  Restricted authnet has remote access by local accounts disabled. $|V|$: number of vertices; $|E|$: number of edges; $\overline{k_{\rm out}}$: average out-degree;  Trees: number of rooted trees (number of sets of reachable systems from $k_{\rm in} = 0$ vertices); Components: number of components; $|RGB|/|GCC|$: size of the residual giant bicomponent relative to the size of the giant connected component, obtained by removing the articulation point with the greatest reduction in size of the giant connected component; $\overline{\epsilon}$: average eccentricity; $d$: graph diameter; $\overline{|V_{{\rm desc}}|}$: average number of descendants of a vertex. Note that the increase in $|V|$ from the original to restricted authnet is an artifact of the clustering, there being more edges and hence more opportunities for vertices that share neighbors to be clustered. Both unclustered networks have $|V| = 3791$.}
\label{laan}
\end{table*}

\subsection{Spreadability and Influence}
For this analysis, we apply a basic SI contagion model to the authnet: a single system is initially infected and then the contagion simply spreads, eventually infecting all reachable systems.  When multiple susceptible systems are reachable from the last infected host, the contagion spreads randomly to any vertex not already part of the infection network.  The model is run until all vertices are infected. The spreading contagion among susceptible hosts is meant to model the essentially unguided lateral movement of the adversary through reachable systems.  The adversary steals credentials and moves among systems using a wide array of techniques, a number of which pertaining to Windows systems are described in Appendix B. We assume that credential theft is always possible and that it is undetected in order to focus solely on the effect of access controls on lateral movement. 
For this simple contagion model, the infection network just corresponds to the Local Administrator authnet itself, with the number of times each host was infected simply given by the in degree of the vertex.  In what follows, we therefore analyze the topology of the authnet to identify important {\it spreaders}, that is, systems that grant or facilitate access to a relatively large segment of the network.   
%Later, when we include the possibility of active sessions (a much more realistic authnet than one considering only local accounts), the important spreaders identified here will be the critical systems linking active sessions.  In short, because they arise from the base Active Directory configuration, they are fundamentally relevant spreaders even as Domain accounts populate the network via active sessions.

Two groups of systems worth considering are those from which the infection spreads to many other systems and those from which the infection can reach far across the network ({\it i.e.} take many hops).  The former are of value to the adversary because, as initial ingress points, they afford a large collection of reachable systems to which access can be expanded.  Meanwhile, the latter might be of interest because a long sequence of accesses enables the adversary to reach points potentially quite unrelated to the initial ingress system.  When we consider active sessions later, exceptional spreaders will be seen to provide routes to privileged sessions; furthermore, unrelated systems tend to contain unrelated data, offering a greater range of exfiltration options to the adversary.  In terms of the authnet, these two groups correspond to vertices with many descendants and those with large {\it eccentricity}.  The eccentricity, $\epsilon$, of a vertex is the maximum-length walk possible from that vertex. 

The number of descendants and eccentricity are good measures of spreadability, particularly when such high-scoring systems are considered ingress points.  But there are systems, perhaps  with relatively few descendants and only average eccentricity, that are nonetheless important for spreadability; for example, systems which sit on the paths between many other systems.  This notion of {\it betweenness} has found several forms in modern network theory: the traditional measure due to Freeman \cite{Freeman} gives high centrality to vertices that tend to lie along the shortest paths between vertices,   
\begin{equation}
\label{bet}
b(v) = \sum_{s,t} \frac{n^v_{st}}{g_{st}} 
\end{equation}
where $n^v_{st}$ is the number of shortest paths between vertices $s$ and $t$ that pass through $v$, and $g_{st}$ is the total number of shortest paths between $s$ and $t$.  For applications to spreadability, however, we should not restrict ourselves only to shortest paths, since contagion will take any route available between hosts. 
%Yet, shortest paths are still the most efficient and direct routes across the network, and so are perhaps relevant to the adversary that knows where it is going.  
A measure that considers all paths between vertices is given by the {\it communicability betweenness} \cite{Estrada},
\begin{equation}  
\omega(v)= \frac{1}{C}\sum_{s,t} \frac{G^v_{st}}{G_{st}},
\end{equation}
where the normalization $C=(n-1)^2-(n-1)$ is the number of terms in the sum, and
\begin{equation}
G_{st} = \frac{1}{\ell !}n_{st} + \sum_{k >\ell}\frac{1}{k!}W^{(k)}_{st}
\label{G}
\end{equation}
where, as in Eq. (\ref{bet}), $n_{st}$ is the number of shortest paths (of length $\ell$) between $s$ and $t$ and $W^{(k)}_{st}$ is the number of paths of length $k$ between $s$ and $t$. The range of $k$ is unbounded.  As above, the superscript ``$v$'' indicates that the paths in question pass through $v$.  The quantity $G_{st}$ is a weighted combination of shortest and longer paths between $s$ and $t$; the weighting adopted in Eq. (\ref{G}), which penalizes longer paths, is that introduced in \cite{Estrada}.   In this analysis, we omit these normalizations so that all paths contribute equally to the betweenness measure.

\begin{figure}[ht]
\centering
\includegraphics[width=0.4\textwidth,clip]{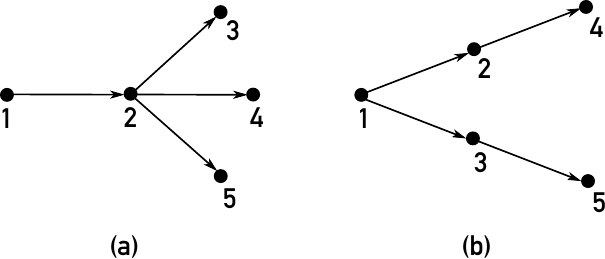}
\caption{Two subgraphs with equal $|V_{\rm desc}|$ and $\epsilon$ but qualitatively different implications for lateral movement, with systems 4 and 5 in (b) possibly more dissimilar than 3, 4, and 5 in (a).  Katz centrality is useful for distinguishing these two cases, and, more generally, is sensitive to how degrees of descendants vary by hop number.}
\label{3}
\end{figure}

Another important aspect of spreadability not yet captured has to do with the number of branchings of the paths descending from a vertex.  For example, consider Figure \ref{3}: the root vertex in these two subgraphs have the same number of descendants and the same eccentricity (and each have zero betweenness). And, yet, they have different implications for the variety of systems reachable as a function of the number of hops from the root node.  For example, the systems reachable in two hops in Figure \ref{3} (a) might be more similar in purpose or data content, given that they are all connected to system 2, than those reachable in two hops in Figure \ref{3} (b), since they do not share a common neighbor.  The {\it eigenvector centrality} \cite{Bonacich} of a vertex takes this branching into account: specifically,
\begin{equation}
e(v) = \frac{1}{\lambda_1}\sum_{s} A_{vs} e(s)
\end{equation}
where $A_{vs}$ is the adjacency matrix of the graph, and $\lambda_1$ is its largest eigenvalue.  Notice that the eigenvector centrality of a vertex can be large if it has many neighbors or if it has (maybe fewer) neighbors with high centrality.  In other words, it's not just about how many friends you have, but also how important they are.  

For directed graphs, the eigenvector centrality is exchanged for the {\it Katz centrality} \cite{Katz},
\begin{equation}
\label{katz}
e_K(v) = \alpha \sum_s A_{vs}e_K(s) + \beta,
\end{equation}
where $\alpha$ acts as an attenuation factor, discounting the importance of more distant vertices, and $\beta$ is the initial centrality given to all vertices irrespective of their degree.  Taking $\alpha = 1/\lambda_1$ minimizes the attenuation, which is desired since spreadability does not depend on the length of associated paths. Typically eigencentralities are bestowed on a vertex by its neighbors, {\it i.e.} a vertex is important if its neighbors are: this is achieved by using the right eigenvector in Eq. (\ref{katz}), which corresponds to incoming edges.  However, in this application we are interested in finding good spreaders, and, hence, we are interested in the reverse interpretation: a vertex has high centrality if it {\it points to} either many other vertices or to those of high centrality.  We therefore use the left eigenvector\footnote{Equivalently, one can employ the right eigenvector on the {\it reversed} graph.} in Eq. (\ref{katz}) with $\beta = 1$.  

Lastly, we consider one additional measure of spreadability, borrowed from the problem of optimal percolation, that is, to identify the minimal set of vertices ensuring the global connectivity of the network.  The vertices that ``glue'' the network together are also the key {\it influencers} with respect to information flow; in our problem, such vertices might be relevant to the lateral spread of the adversary across well-connected regions of the network.  The {\it collective influence} of a vertex at level $\ell$ is defined \cite{Morone},
\begin{equation}
\label{CI}
{\rm CI}_\ell(v) = (k(v) - 1) \sum_{s \in \partial {\rm Ball}(v,\ell)} (k(s) - 1)
\end{equation}
where $k$ is the degree centrality and $\partial {\rm Ball}(v,\ell)$ is the frontier of the ball of radius $\ell$ defined around vertex $v$. If in Eq. (\ref{CI}) we use $k_{\rm out}(v)$, the out-degree of vertex $v$,  the collective influence at the level $\ell$ characterizes the ability of the vertex $v$ to influence the spread of contagion to systems $\ell$ steps away.  Interestingly, this measure is ambivalent to the structure of the graph in between $v$ and the level-$\ell$ systems. As with $e_K$, collective influence should not discount long paths, and so we take $\ell$ to correspond to the network diameter.  

\begin{figure}[ht]
\centering
\includegraphics[width=0.7\textwidth,clip]{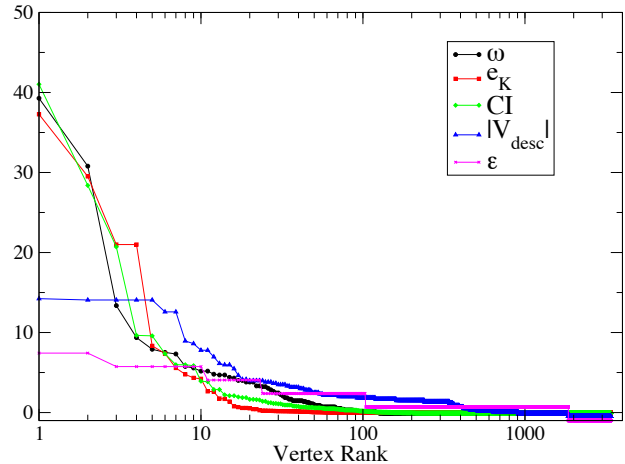}
\caption{The five (standardized) spreadability measures discussed in the text.  The vertices are ranked within each measure separately. Communicability betweenness ($\omega$), Katz centrality ($e_K$), and collective influence (CI) have clear outliers, whereas the distributions of the number of descendants ($|V_{\rm desc}|$) and eccentricity ($\epsilon$) have less extreme values.}
\label{4}
\end{figure}

To summarize, we consider five metrics of spreadability: $(|V_{\rm desc}|,\epsilon,\omega,e_K,{\rm CI}_\ell)$.  Although there is sure to be some overlap and redundancy in this set, each measure seeks to capture a unique feature of the network.  The values of each measure for each vertex are presented in Figure \ref{4}: here, we rank each vertex within each measure and plot the rank of the vertex along the $x$-axis.  The measures have been standardized for direct comparison: notice that $\omega$, $e_K$, and ${\rm CI}_\ell$ each have clear outliers, whereas $|V_{\rm desc}|$ and particularly $\epsilon$ have less extreme values.  Rather than focus on any single metric, we consider all of them together and look for outliers in the full five-dimensional parameter space.

The distribution in the full five-dimensional measure space is not fit well by the most common parametric distributions---it is highly concentrated at low centrality with sparse outliers, suggesting the use of density-based clustering to mine for outliers\footnote{The distribution of centrality measures likely depends sensitively on the nature of the network and so could conform to a parametric distribution in certain applications.  In those instances, outliers should be found with respect to the particular distribution ({\it e.g.} the generalized extreme Studentized deviate works well for Gaussian-distributed data in the presence of multiple outliers.)}.  Specifically, the DBSCAN (density-based spatial clustering of applications with noise) algorithm \cite{DBSCAN} is used to identify points in the rarefied regions of the centrality parameter space, far from the bulk of low-centrality vertices.  DBSCAN has two tunable parameters: the minimum number of points, {\it MinPts}, required to form a cluster, and $\varepsilon$, the maximum distance that a point may lie from a cluster to be considered part of that cluster.  In practice, usually {\it MinPts} is chosen and $\varepsilon$ is determined by finding the ``elbow'' in the plot of the distance from each point, ordered from smallest to largest, to its {\it MinPts}-nearest neighbor.  Any point that lies farther than a distance $\varepsilon$ from its nearest cluster is considered by DBSCAN to an outlier.  With {\it MinPts} $= 20$ we identify 56 outliers\footnote{There is no ``right'' value for {\it MinPts}; its value depends on the number of desired outliers, which should be based on the exposure risk tolerance.} out of 3389 systems, or roughly 1.5\%.  

\begin{figure}[ht]
\centering
\includegraphics[width=0.7\textwidth,clip]{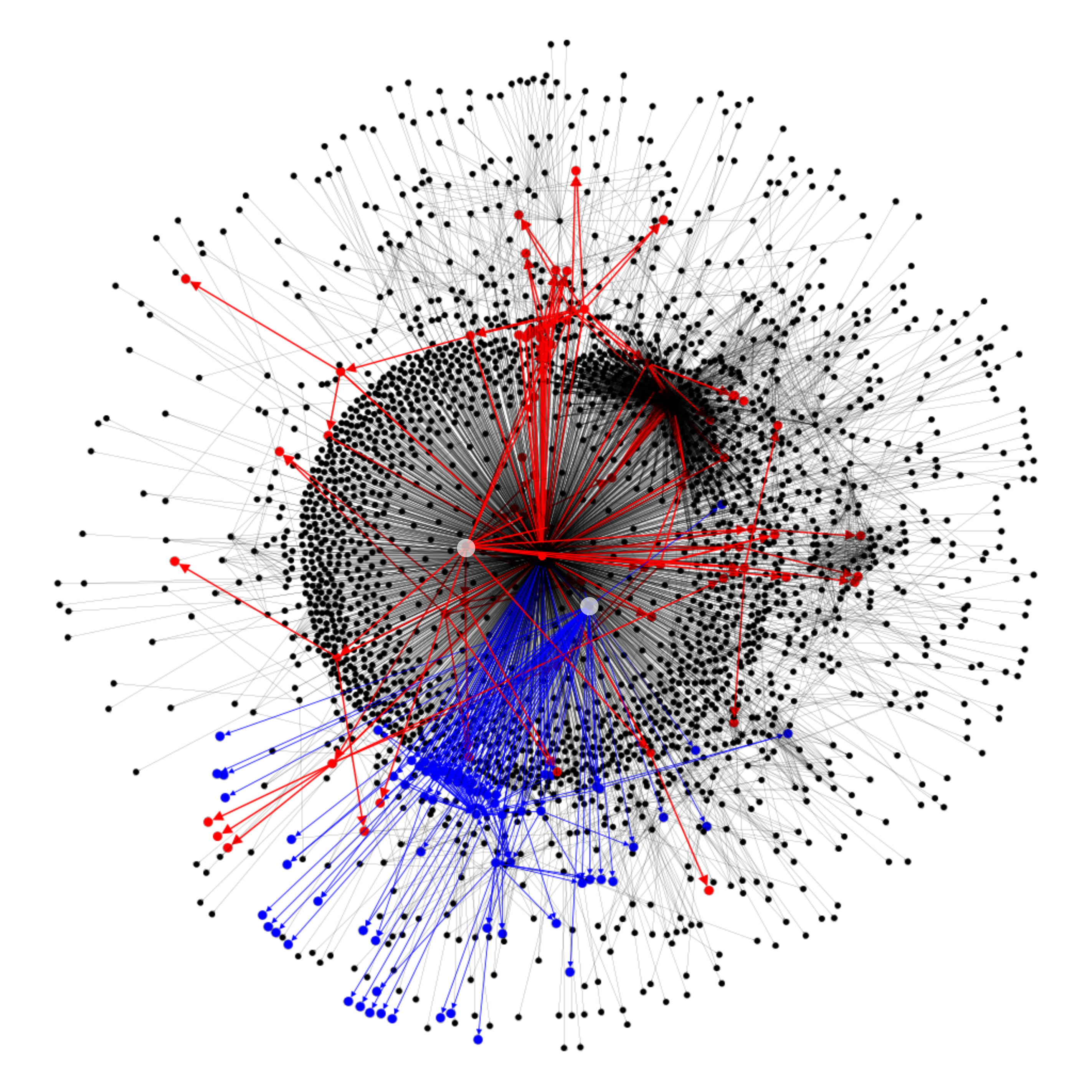}
\caption{Descendants of the top collective influencer (blue paths) and most Katz-central system (red paths). See text for discussion.}
\label{5}
\end{figure}
The spreadability of two prominent outliers is displayed in Figure \ref{5}.  The blue paths trace the descendants of the top collective influencer, and the red paths trace the descendants of the system with largest Katz centrality.  Interestingly, $|V_{\rm desc}| = 80$ for both systems, though the two edge sets are mostly disjoint.  The top collective influencer has a rather large out-degree, $k_{\rm out} = 73$, and small eccentricity, $\epsilon =  2$, compared to the most Katz-central system, with $k_{\rm out} = 26$ and $\epsilon = 5$; as a result, the distributions of descendants across the network are quite distinct.  

Before moving on, it is of interest to compare this approach with $k$-core analysis, which is based on the premise that the most efficient spreaders in complex networks are those vertices within the core of the network \cite{Kitsak}.  These methods apply best to undirected networks, in which the contagion can spread between any vertices sharing an edge; here, however, the directed nature of the epidemic makes $k$-core analysis less useful.  For directed graphs, a vertex belongs to the ``out'' $k_S$-core if it belongs to the maximal subgraph in which every vertex has $k_{\rm out} \geq k_S$.  The $k_S$-shell is the collection of vertices that belong to the $k_S$-core but not the $(k_S+1)$-core.  In \cite{Kitsak}, it is shown that vertices within the highest $k_S$-shell are particularly efficient spreaders, however, due to the heavily fragmented nature of the directed authnet, the largest $k$-core has only $k_S=3$, a set of four vertices which does not include any of our influential spreaders.  In fact, all but one of the influential spreaders are in the $1$-shell, shared by around eighty other vertices.   
\subsection{Countermeasures}
To reduce spreadability, a variety of countermeasure and mitigations are available.  We consider two in this analysis: one that targets specific spreaders, modifying their access controls to reduce their influence; and one that acts globally, by disabling remote access for all local accounts.
\subsubsection{Tightening Access Controls} 
The presence of outliers, particularly those at the extremes of the five-dimensional centrality distribution, might signal an access control problem.  In the case of the top collective influencer, the broad and direct access to a large set of systems might be examined: perhaps a network rather than local account would reduce the exposure of this account to compromise, since credentials would only be available on the system during an active session.  In the case of the most Katz-central system, its immediate direct access is not especially overwhelming: it is the far reach of this account, via credential chaining, that is a potential problem.  The trouble is thus not with any individual account, but rather with the whole sequence of connected accounts taken together. 
\begin{figure}[ht]
\centering
\includegraphics[width=0.5\textwidth,clip]{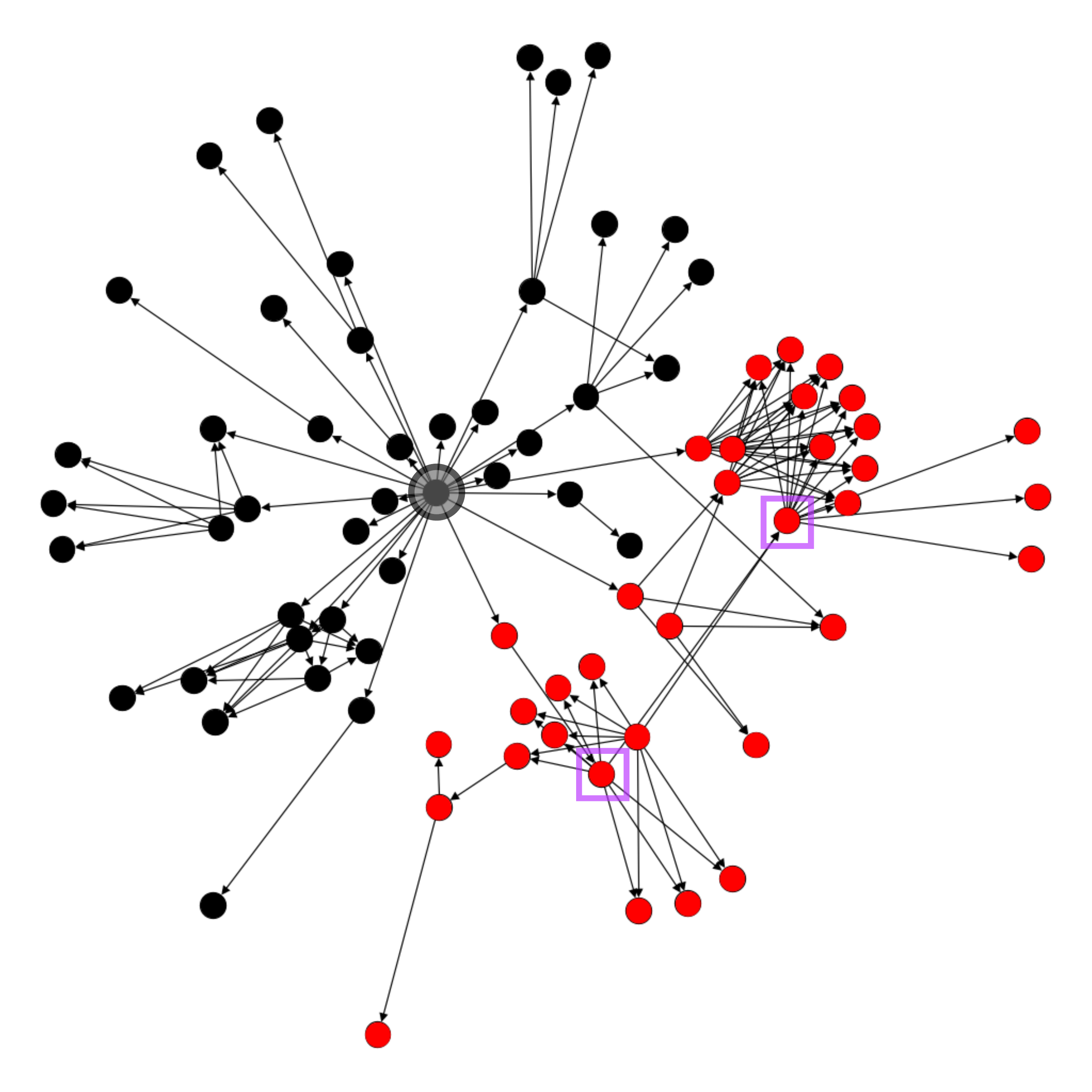}
\caption{The descendants graph, $G(E_{\rm desc},V_{\rm desc})$, of the most Katz-central system of Figure \ref{5} (black disc).  Vertices are colored according to Girvan-Newman community detection.  The two systems with largest betweenness are highlighted by the purple squares.}
\label{6}
\end{figure}

There are a range of techniques one might employ to analyze a given system's descendants graph, $G(E_{\rm desc},V_{\rm desc})$. For example, identifying community structure might suggest edges or vertices whose removal would reduce the centrality of the spreader system.  The removal of edges corresponds to the elimination of local account access to the destination systems, whereas the removal of vertices corresponds to either disabling the account or the elimination of all access.  As an example, suppose we wish to eliminate select accesses to reduce the spreadability of a system:  in Figure \ref{6}, the Girvan-Newman method \cite{Girvan} is applied to resolve $G(E_{\rm desc},V_{\rm desc})$ into two communities by minimizing the number of edges between them.   Removal of the edges that bridge the two communities, corresponding in this case to the elimination of three accesses, would help reduce the centrality of the spreader system.  This method can be iterated to find ever-smaller communities, incrementally breaking long credentials chains.  Alternatively, one can look for vertices with high betweenness (or any of the other centrality measures discussed above) and examine the access controls of accounts that authenticate to those systems, and/or the access controls of the local accounts on those systems.  

There are many approaches to the problem of finding ``linchpin'' systems in graphs, including efficient articulation point removal \cite{Tian}, graph fragmentation \cite{Borgatti}, equal graph partitioning \cite{Chen}, and other means of community detection, and our intent is not to study each in depth.  The purpose of this section is merely to illustrate how this remedial analysis might be done in the face of trouble accounts identified via the preceding outlier analysis.  
\begin{figure}[ht]
\centering
\includegraphics[width=0.7\textwidth,clip]{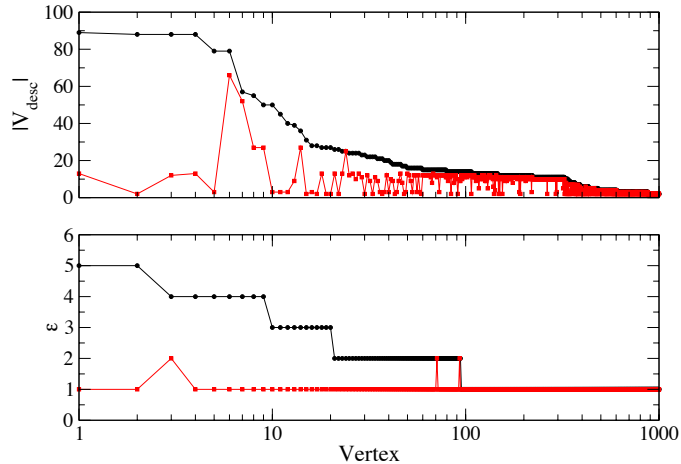}
\caption{Comparison of $|V_{\rm desc}|$ and $\epsilon$ between networks with and without remote access for local accounts disabled. Vertices are ordered according to their values of  $|V_{\rm desc}|$ and $\epsilon$ for the unrestricted network (black), with red squares indicating the value of the vertex in the restricted network.}
\label{7}
\end{figure}
\begin{figure}[ht]
\centering
\includegraphics[width=0.7\textwidth,clip]{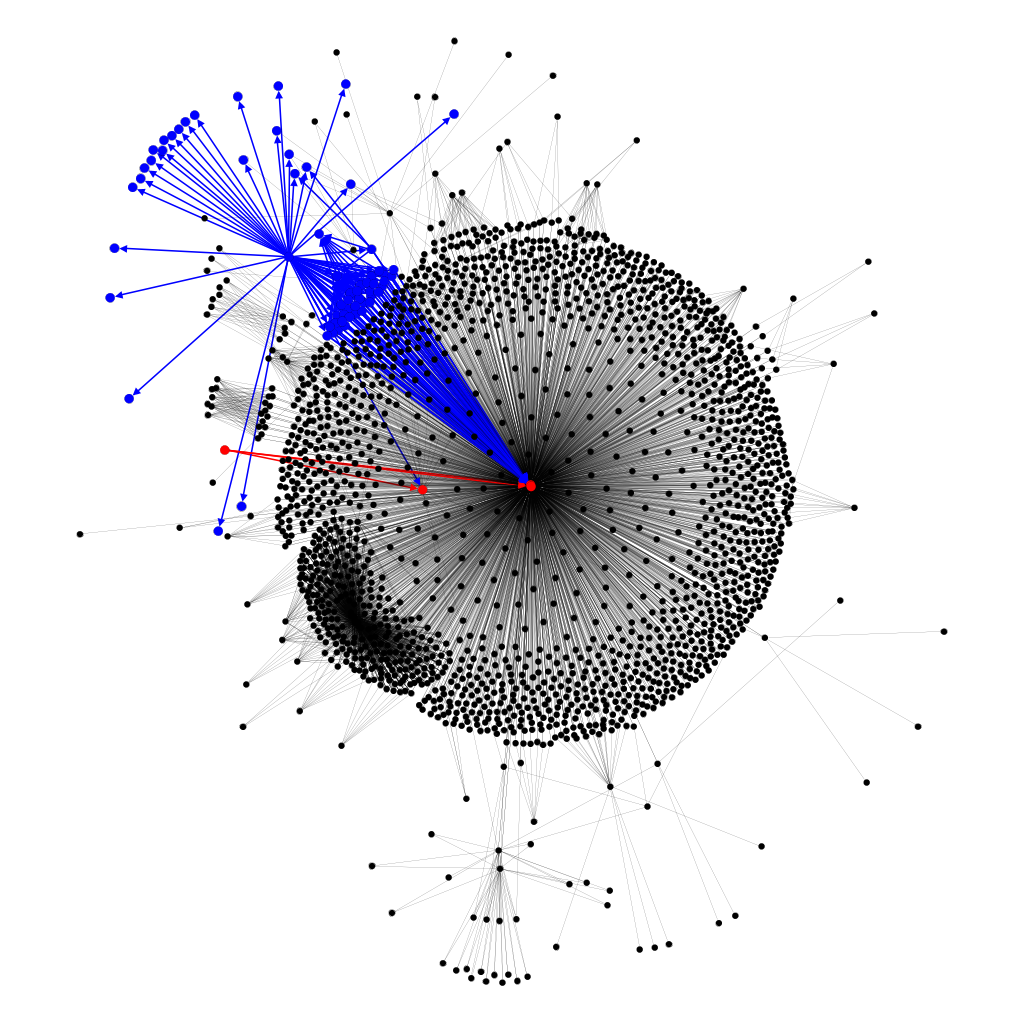}
\caption{Descendants in the restricted authnet of the top collective influencer (blue paths) and most Katz-central system (red paths) from analysis of the original authnet, Figure \ref{5}.}
\label{8}
\end{figure}
\subsubsection{Disabling Remote Access for Local Accounts}
We next study what happens to spreadability when remote access is disabled for all local accounts. The authnet corresponding to this restriction can be obtained by pruning all ``Admin to'' edges directed away from User-type vertices in Figure \ref{1}, leaving only group-delegated Administrator relationships.  The resulting authnet is reduced in size, as expected, with properties detailed in Table 1.  Interestingly, there is a dramatic effect on spreadability: comparison of $|V_{\rm desc}|$ and $\epsilon$ of the top vertices from the unrestricted authnet with those of the restricted authnet shows that the number of descendants of top spreaders are greatly reduced, Figure \ref{7} (top), as well as their eccentricity, Figure \ref{7} (bottom).  In general, this restriction on remote access breaks long credential chains.  Recalling the top collective influencer and most Katz-central system from earlier (Figure \ref{5}), the effect on the credential chains emanating from these systems is shown in the restricted authnet, Figure \ref{8}.  The most Katz-central system which had $|V_{\rm desc}| = 80$ and $\epsilon = 5$ with unconstrained remote access is reduced to $|V_{\rm desc}| = 3$ and $\epsilon = 1$ when remote access is disabled for local accounts; meanwhile,  the top collective influencer with $|V_{\rm desc}| = 80$ and $\epsilon = 2$ is more modestly affected, reduced to $|V_{\rm desc}| = 67$ with no change in eccentricity.   Indeed, there are still outliers:  DBSCAN identifies 13 outliers in the five-dimensional metric space, but these outliers are markedly less extreme, as Figure \ref{7} suggests\footnote{We only compare across the measures $|V_{\rm desc}|$ and $\epsilon$ rather than the full set of five because the others are normalized to the size of the network, making a comparison across networks difficult.}.   
%Before moving on, it must be mentioned that the most effective protection against lateral movement via local accounts is to simply disable remote access for local accounts.  This option has been available as a Local Policy since Windows 8.1 and Windows Server 2012; however, Domain-wide implementation might not be possible on all networks.     I don't think this is actually relevant here: while we are talking about local accounts on system A authenticating to system B, this access is primarily via group membership at the Domain level and not local policy.  Perhaps try it with only group membership links and see what happens?
\subsection{Spreaders as Ingress Points}
While we have identified a collection of systems with an unusual ability to spread access across the network, it is important to quantify the expectation that an adversary might find themselves on such a system in the first place.  Our outlier analysis found that around 1.5\% of systems were notable spreaders, and so under the naive assumption that the adversary penetrates the network at random with an equal chance of exploiting any given system, there is a likewise 1.5\% chance that the adversary ingresses the network on a spreader.  But this assumption is obviously poor, since certainly at least {\it some} of the systems on the network are unreachable from the outside.  Consider instead the more realistic case in which the adversary achieves ingress via email phishing, a prevalent attack vector involved in 70\% of breaches associated with nation-states or state-affiliated actors in 2017 \cite{Verizon2018}. Here, the target systems are likely workstations, of which there are close to 2000 on the network.  Interestingly, all of the spreaders identified in the preceding analysis are also workstations, and so the chance of randomly ingressing the network via a spreader system is doubled, to around 3\%.  Alternatively, had some of our spreaders been outward-facing servers, we might instead conduct a risk analysis of the exploitability of such systems across the network.  The point here is not the specific numbers, but that exposure risk depends on the make-up of the network, the outliers, and the intrusion model assumed.
\section{Finding Escalators}
The foregoing analysis addressed the authentication relationships among Local Administrator accounts across the network.  This is useful and sufficient for finding prominent spreader systems resulting from access control configurations.  It is common, however, to find Administrator accounts that are not individually local to a given system, but instead are network-level accounts controlled via group policy.  Administrators in different network groups generally have different levels of access and privilege, called {\it security tiers}.  The Microsoft Active Directory Administrative Tier Model \cite{MADATM}, for example, separates the enterprise into three security tiers: Domain Administrators occupy the top tier and generically have the broadest access and greatest privilege, with the ability to access and modify data on any Domain system, including Domain Controllers.  Lower-tiered groups include server and workstation Administrators.  

From the perspective of credential residue, the difference between local and network-level Administrators is that the latter credentials are generally stored statically and indefinitely only in the centralized database (the Domain Controller on Windows networks), and so cannot be extracted from the local security databases of individual systems to which they have access.  But, these credentials can still possibly be found in the running memory or local cache of systems on which these accounts have established active sessions.  Privilege escalation from Local Administrator to a network-level Administrator, or from a lower- to higher-tier network-level Administrator, therefore generally requires the compromise of active sessions.  For example,  an adversary might escalate privileges by compromising a Desktop Administrator's account and locating the active session of a Windows Domain Administrator on a workstation in the lesser Administrator's tier.  Lesser Administrators should therefore never be permitted to access an asset at a higher tier, and higher-tier Administrators should limit access to lower-tier assets.  Of course, it is necessary for Domain Administrators to be occasionally allowed access to workstations, and so this is not an access control issue {\it per se}; rather, it is one of active session policy, of who can log in where and when.

\begin{figure*}[ht]
\centering
\includegraphics[width=0.85\textwidth,clip]{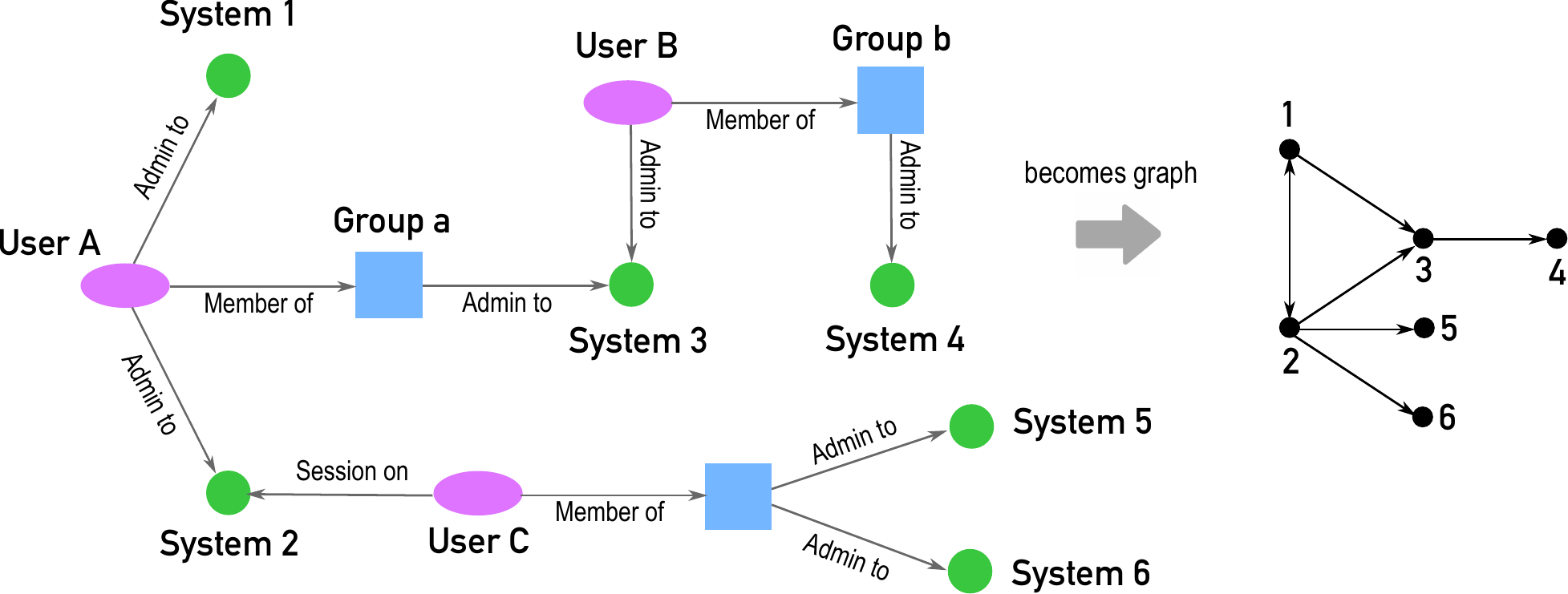}
\caption{Illustration of how active sessions modify the authnet. User C has an active session on System 2; if User C's credentials are available there, then System 2 in the authnet now has authentication relationships with all the systems that User C has Administrative access to.}
\label{19}
\end{figure*}

Using data on current and recent active sessions, we create a graph of authentication relationships made possible by the credential residue left by these sessions, and combine it with the Local Administrator authnet to obtain a more comprehensively and timely map of authenticated routes across the network.  If a user's credential is available on a system because of an active session, the combined authnet is constructed as shown in Figure \ref{19}.  This is an extended version of the Figure \ref{1}.  The vertex with the active session ``sprouts'' edges to all systems that the user has privilged access to.   We then search this {\it combined authnet} for pairs of systems, one hop apart, which have concurrent active sessions by accounts with different amounts of access, {\it i.e.} the two sessions belong to two different security tiers.  We refer to the lower-tier system of such pairs as an {\it escalator}.  Due to credential chaining, compromise of the escalator effectively grants the adversary with the access of the higher-privileged session because it is only one credential theft away.  Escalators embody the Microsoft concept of cross-tier logins, in which accounts in different security tiers access one another's assets.  

\subsection{Combined Authentication Network}
On Windows networks, the primary source of active session data are server and client Windows Event Logs.  Specifically, we extracted all Administrative logon (Event Code 4624), log off (4634 and 4647), and system reboot events (6005) over the course of seven days\footnote{Ideally, the historic record should span a time period longer than the duration of all active sessions and, conservatively, the uptimes of systems with sessions in the record.} to establish a current record of active sessions.  To create an authent from this session data, it is necessary to understand where the credentials are: this depends on the type of logon (interective, network, or remote interactive), authentication package (NTLMv2 or Kerberos), impersonation level, as well as any protections that are in place, like Protected Users groups \cite{Plett}, Restricted Admin Mode \cite{Falde}, Remote Credential Guard \cite{Lich}, or a ``no caching'' policy for Domain accounts on the destination host.  Table A.3 in Appendix A summarizes the credential residue expected given different combinations of these factors.

In total, active sessions with credential residue are found on 362 systems, or just around 10\% of all systems. Since these sessions belong to Administrators in Domain groups with generally wide access, systems with active sessions will tend to have very high out-degree.  This means that there are many more edges in the combined authnet: 547,334 edges to the Local Administrator authnet's 8747 (cf. Tables 1 and 3).  A depiction of the combined authnet, like Figure 2 for the Local Administrator authnet, is therefore not legible and so we do not present one.  
%The transitivity, $C$, of the combined authnet is small compared to the Local Administrator authnet: $C$ is defined as the number of closed triplets (three vertices with three edges between them) relative to the number of all triplets, which includes open triplets (three vertices with two edges between them).   Because of the great increase in $\overline{k_{\rm out}}$, there many open triplets formed by vertices with active sessions since these accounts have many accesses (making such vertices look like hubs).  Also related to this great increase in connectivity across the network is a reduction in number of rooted trees (vertices can reach more descendants), and fewer components (singleton systems can be accessed by accounts with active sessions).
\subsection{Escalators and Associated Accounts}
\begin{figure}[htp]
\centering
\includegraphics[width=0.5\textwidth,clip]{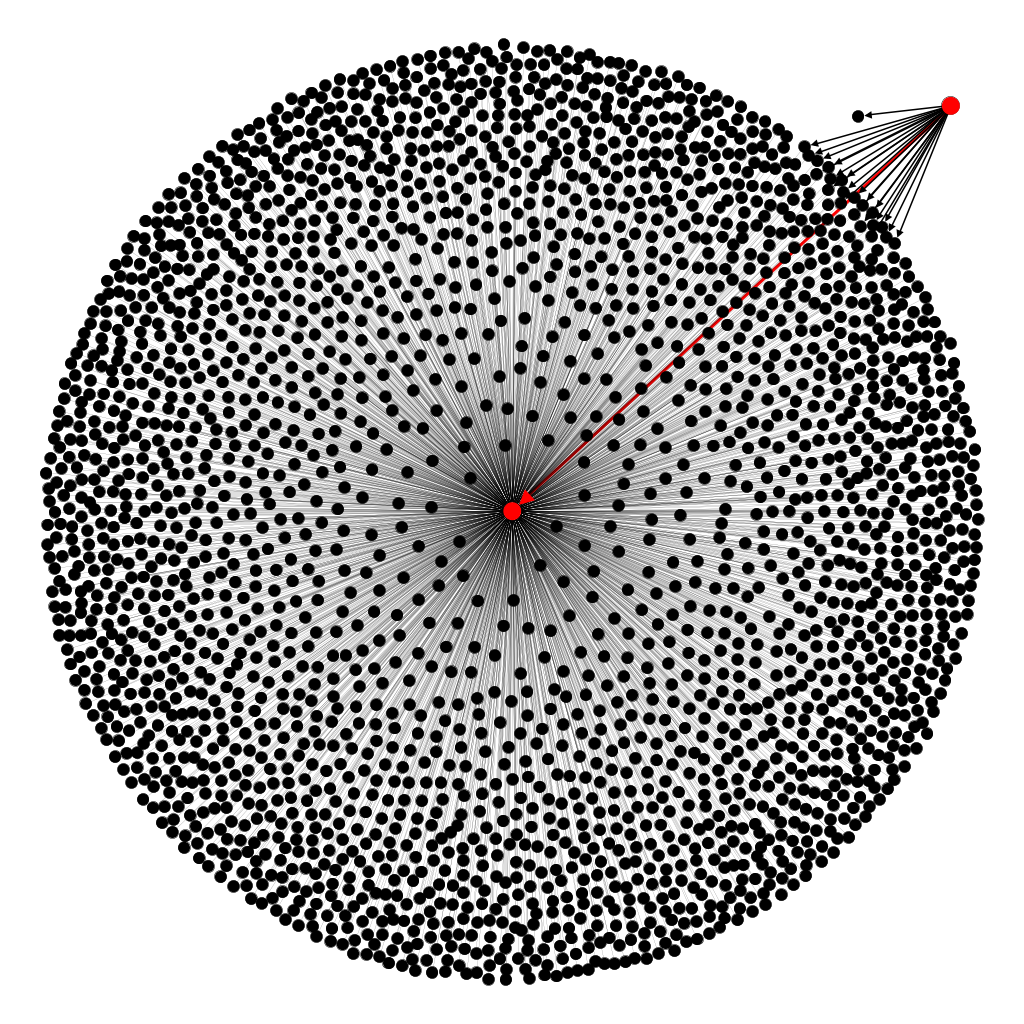}
\caption{Subgraph indicating a substantial privilege escalation: the administrator with an active session on the smaller out-degree system ($k_{\rm out} = 22$, red vertex in upper right) can access a system on which an administrator with higher privilege (greater access) has an active session ($k_{\rm out} = 2450$, red vertex in center of large cluster). For this escalator, $\Delta |N(u)|_{\rm max} = 2429$ and $\overline {\Delta |N(u)|}=110$.}
\label{9}
\end{figure}

Within the combined authnet, escalators appear as in Figure \ref{9}.  To identify important escalators, it is useful to consider two different contagion strategies.  One is a targeted escalation, in which the adversary has knowledge of where privileged active sessions are relative to their location.  Such knowledge could be acquired with network and account reconnaissance, of the type achieved with tools like Bloodhound.  Under the assumption that the adversary would select the largest escalation under this strategy, escalators can be quantitatively ranked according to the maximum increase in the size of the neighborhood surrounding the vertex with the high-tier session, $v$, relative to the neighborhood around the escalator, $u$, or 
\begin{equation}
\label{maxN}
\Delta |N(u)|_{\rm max} = \max_v |N(v) - N(v)\cap N(u)|.
\end{equation}
The {\it neighborhood} of a vertex, $u$, is the set of vertices reachable from $u$ in one hop.  

In the alternative strategy, the adversary might have little-to-no knowledge of privileged sessions, and instead seeks to escalate privileges by moving randomly from their current location.  In this case, a relevant metric is the increase in neighborhood size averaged over the out degree of the escalator,
\begin{equation}
\label{avgN}
\overline {\Delta |N(u)|} = \frac{1}{k_{\rm out}(u)}\sum_{v \in N(u)} |N(v) - N(v)\cap N(u)|.\\
\end{equation}

\begin{figure}[ht]
\centering
\includegraphics[width=0.7\textwidth,clip]{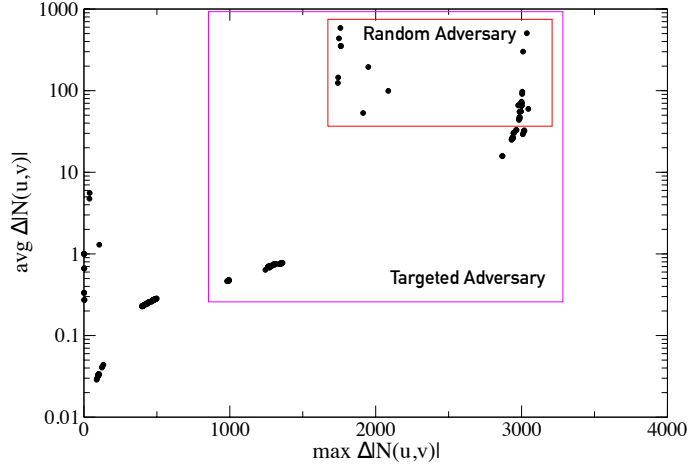}
\caption{$\overline{\Delta |N(u,v)|}$-vs-$\Delta|N(u,v)|_{\rm max}$ plane showing escalators.  Those with large $\overline{\Delta |N(u,v)|}$ offer substantial privilege escalation to an adversary that chooses its next hop randomly, whereas those escalators with large $\Delta|N(u,v)|_{\rm max}$ but small $\overline{\Delta |N(u,v)|}$  would be most beneficial to a targeted adversary that knows where the privileged sessions are.}
\label{10}
\end{figure}

The combined authnet was analyzed and 348 escalators were identified connecting to 295 higher-privileged sessions, with a total of 362 systems having active sessions. The escalators were evaluated according to Eqs (\ref{maxN}-\ref{avgN}), Figure \ref{10}. Unlike the spreader analysis, an outlier assessment isn't appropriate because most of the escalators have significant $\Delta|N(u,v)|_{\rm max}$ and $\overline{\Delta |N(u,v)|}$.  Indeed, many escalators facilitate access to over 1000 new systems, and a smaller group have an exceptionally high average number of new accesses per neighbor.  These results signal a  problem with cross-tier sessions, but how big is the risk?  That is, how likely is the adversary to find either an escalator or a high-privilege session on the network? 
\begin{figure}[ht]
\centering
\includegraphics[width=0.7\textwidth,clip]{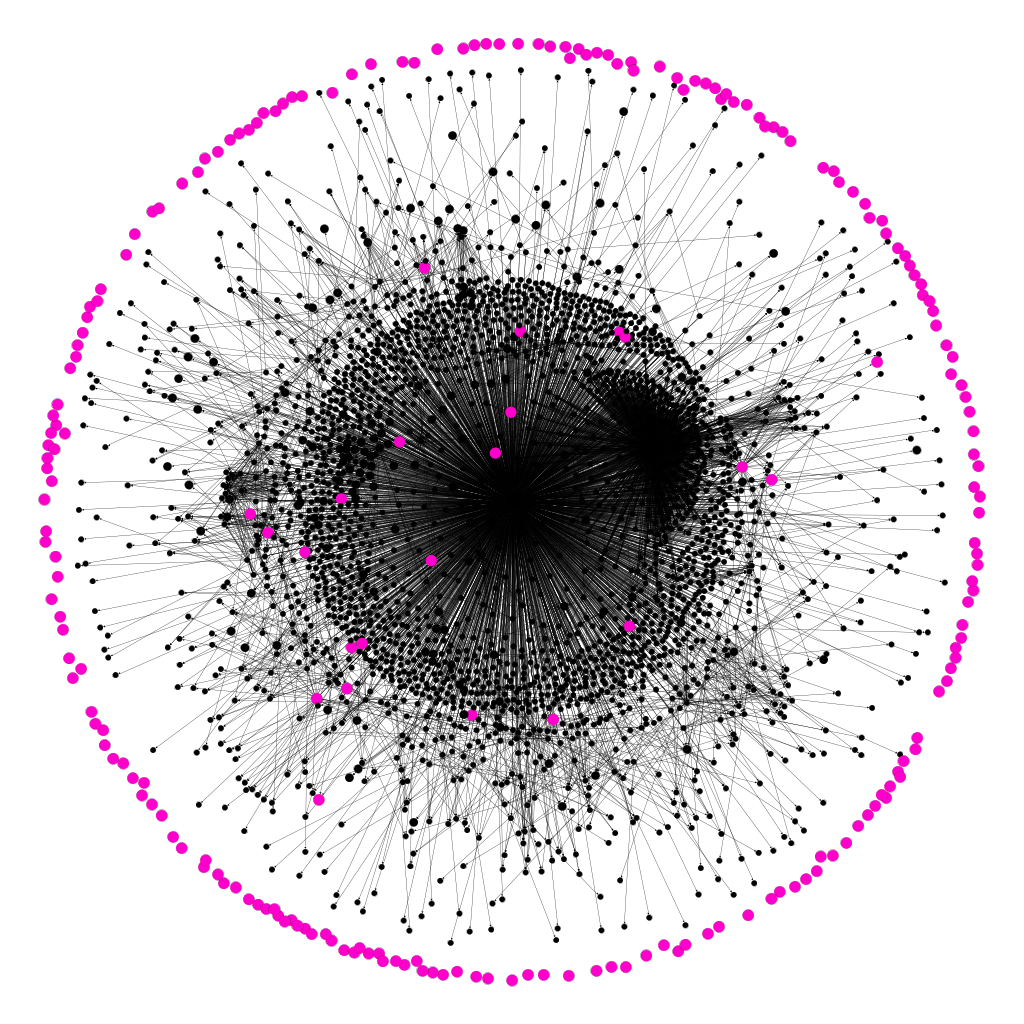}
\caption{Systems with active sessions overlayed on the Local Administrator authnet.  Many of them are on isolated vertices unreachable by other local accounts.}
\label{17}
\end{figure}

\begin{figure*}[ht]
\centering
\includegraphics[width=0.75\textwidth,clip]{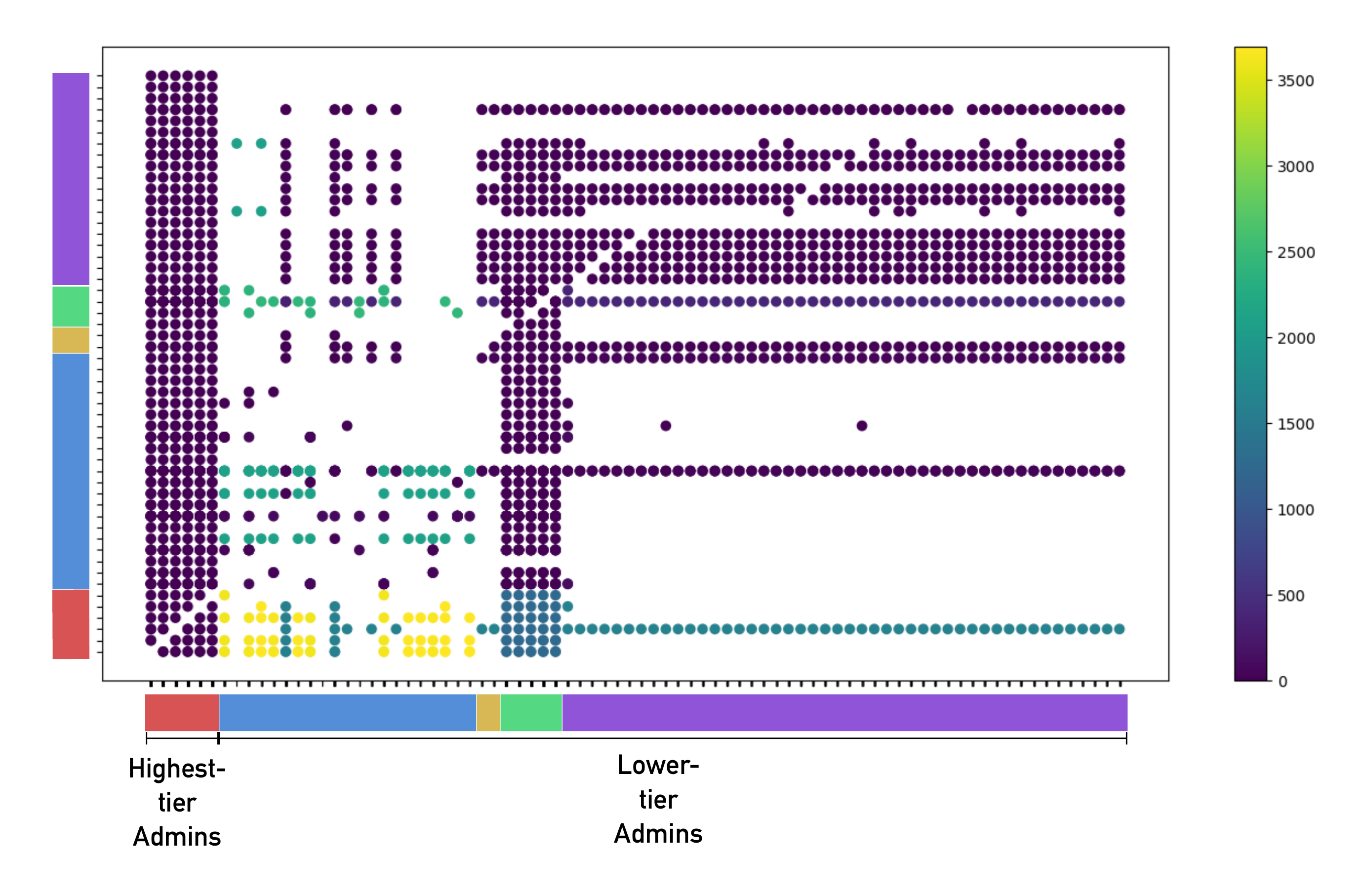}
\caption{Privilege escalation exposure by account.  Colors indicate number of new accesses granted through successful privilege escalation from accounts along the $x$=-axis to those along the $y$-axis.}
\label{11}
\end{figure*}
Suppose the adversary has obtained Local Administrator access to some workstation and is seeking to escalate privileges to an Administrator in a Domain group with broader access.  Thirty-five of the approximately 2000 Local Administrator accounts can reach at least one active session (Figure \ref{17}).   Only 23 distinct sessions of 362 can be reached this way, indicating that privilege escalation from Local Administrator is difficult to do (2\% of Local Administrators can reach 6\% of sessions)---this isolation of privileged sessions is desirable, suggesting good local account access configurations.  Alternatively, we can ask how easy it might be for the adversary to escalate to the highest security tier, like an account in the Domain Administrator's group, in one hop.  There are 35 active Domain Administrator sessions in our data set, and 149  lower-tier sessions are escalators with respect to these accounts.  Suppose that these escalator systems are primarily at risk of compromise by remote exploit, say, via phishing email as before.  Of the 149 sessions, 78 of them are workstations assumed susceptible to phishing.  If the adversary targets the 1877 workstations at random, there is thus a 4\% chance of phishing one that can immediately escalate to Domain Administrator.  This is all just to demonstrate the kinds of exposure analysis possible with this approach. 

Interestingly, of the 35 workstations that can reach active sessions, 14 are spreaders, eight of which can reach two or more escalators.  And so, while the spreaders were found using generic graph centrality measures applied to a basic contagion model, they are also key to lateral movement targeting specific systems with privileged active sessions.

\subsection{Countermeasures}
Reducing privilege escalation exposure is a general problem with both targeted and network-wide countermeasures.  One targeted approach involves identifying specific privileged accounts whose cross-tier logins have greatest privilege escalation exposure: if the account must access these lower-tiered systems, a ``staggered'' session policy could be adopted to ensure that accounts in different security tiers are never simultaneously one hop away from each other.  Alternatively, on Windows networks there are a variety of global countermeasures that could be implemented across the Domain to reduce the credential residue on systems.  {\it Restricted Admin Mode} is a feature available since Windows 7 / 2008 R2 that authenticates users over remote interactive logon as Local Administrators.  Doing so obviates the need for users to send credentials to the remote system, but since the session belongs to the remote system's Local Administrator, delegation is broken and additional remote services cannot be accessed on the authenticated user's behalf (see Appendix A.3). An improvement that restores delegation while keeping credentials safe is offered by {\it Windows Remote Credential Guard}.  This feature authenticates Type 10 logins over Kerberos and places artifacts on the remote host in a special, virtualized portion of LSASS memory which protects them from kernel processes.  When additional resources are needed, the remote host directs the request to the client which has access to the virtualized container so that delegation can proceed.  Finally, users in the {\it Protected Users Group}, available since Windows 8.1 / 2012 R2, are forced to authenticate over Kerberos and won't have plaintext credentials or Kerberos artifacts stored in LSASS, breaking delegation\footnote{Despite these advancements in credential protection, nothing is ever perfect.  For example, recent versions of \texttt{mimikatz} are able to snag credentials on a Remote Credential Guard-protected system before they end up in protected storage by injecting a custom SSP into LSASS that stores them in regular memory outside the virtualized container.  Alternatively, as long as a user has an active session on a remote system, whether credentials are stored or not, their access token resides in memory and can be hijacked; with the token, an attacker can request services on the user's behalf and the client will be tricked into providing service tickets.}.

We first examine how to detect accounts that frequently cross security tiers, and then we use the authnet to test the effectiveness of implementing Windows Remote Credential Guard on all systems. 
\subsubsection{Eliminating Cross-Tier Logins}
To reduce the significance and existence of escalators in the network cross-tier logins must be minimized.  It is therefore helpful to identify the specific Administrator accounts involved in the active sessions.  Figure \ref{11} is a chart depicting the privilege escalation exposure of all accounts with active sessions.  
All accounts with active sessions are given along the $x$-axis (each tick mark is one account).  For a given account along the $x$-axis, the dots mark those other accounts along the $y$-axis that have sessions within its neighborhood.  The coloration indicates how many new systems are accessible via privilege escalation from the $x$-axis account to the $y$-axis account.  

Accounts are organized broadly into ``highest-tier'' and ``lower-tier'' sets, with the lower-tier accounts belonging to several distinct groups (like desktop, helpdesk, and server Administrators). This visualization enables us to easily verify that almost all of the high-tier Administrators login to systems below their security tier, most often accessing systems also accessible to the ``blue'' Administrators.  And, such logins carry significant liability: if one of the ``blue'' accounts is compromised, privilege escalation to the highest tier is possible with an increase in access of some 3000 systems.  The chart also reveals some good practices: most Administrator groups avoid the ``purple'' Administrator's tier, and, aside from the ``blue'' group, accounts within the same tier have approximately the same access (little-to-no increase in access in moving from one account to another in the same group).   
\subsubsection{Reducing Credential Residue}
Interactive sessions are especially worrisome from the perspective of credential residue since impersonation and delegation require that credentials be placed into LSASS on the client and server (for remote logins)---in our data sample, around 70\% of systems with active sessions had at least one Type 10 login.   The protective features discussed in Section 4.3 aim to reduce or eliminate the credential residue on remote systems left by interactive logins.  To demonstrate our methodology's ability to test out different countermeasures, we now examine what effect Windows Remote Credential Guard has on privilege escalation exposure. 

\begin{figure}[htp]
\centering
\includegraphics[width=0.7\textwidth,clip]{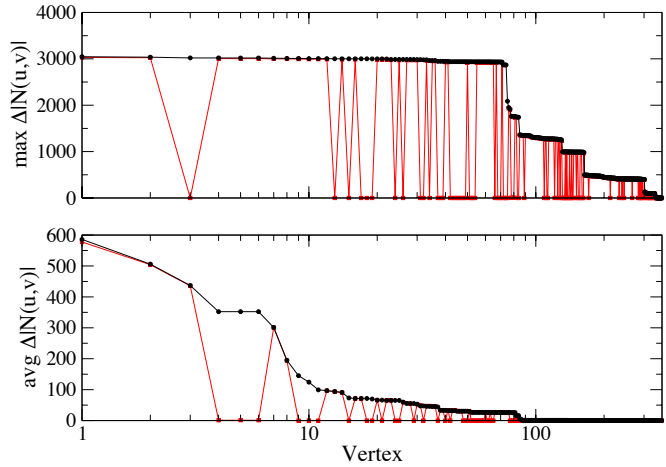}
\caption{Comparison of $\Delta|N(u,v)|_{\rm max}$ and $\overline{\Delta |N(u,v)|}$ between escalators from the unprotected network (black) and one with Windows Remote Credential Guard in place on every system (red). Vertices are ordered according to their values of these metrics in the unprotected network  (black), with red squares indicating the value of the vertex in the protected network.}
\label{12}
\end{figure}
A network with Remote Credential Guard has a different combined authnet than one without; to build it, we use the rules shown in Table A.3.  The resulting graph is markedly different than the original active session authnet, Table 3.  In summary, with Remote Credential Guard in place on all systems in the Domain the number of escalators is reduced from 348 to 262, a reduction of around 25\%.  Figure \ref{12} shows that many of the eliminated escalators are indeed significant, with around one third of escalators with $\Delta|N(u,v)|_{\rm max} \gtrsim 2800$ eliminated by the Credential Guard.  This countermeasure also has implications for the exposure of the highest-privileged accounts; the 35 Domain Administrator sessions found to have credential residue on the unprotected network are reduced to 25, with now only 94 escalators (versus 149) able to reach these sessions, a 37\% reduction.  Interestingly, though, the systems most affected by the Remote Credential Guard tended to be servers, and so the risk of compromise of an escalator via phishing a workstation is unchanged. 

%Notice that Figure \ref{12}  also reveals some escalators that become {\it more} significant after Remote Credential Guard is emplaced.  In these cases, there were multiple sessions by accounts with varying degrees of access on each system; Remote Credential Guard eliminated the most privileged of these, leaving only the lower-privileged sessions.  The higher-privileged sessions that these escalators could access remained unaffected by the mitigation, and so as a result what was a moderate gain in access in going from an already highly-privileged account to another high-privilege account, becomes a large gain in access in going from a lower-privilege account to this same high-privilege account.

\begin{table*}[htp]
\footnotesize
\begin{center}
\begin{tabular}{|c|c|c|c|c|c|c|c|c|c|}
\hline
authnet&$|V|$&$|E|$&$\overline{k_{\rm out}}$&Trees&Components&$|RGB|/|GCC|$&$\overline{\epsilon}$&$d$&$\overline{|V_{\rm desc}|}$   \\ \hline
combined&3601&547,334&152&77&55&0.98&1.02&7&373 \\
\hline
%\vtop{\hbox{\strut local account}\hbox{\strut + sessions}}&3716&280617&132&0.007&0.45&133&56&0.97&0.7&7&159&145 \\
combined&\multirow{2}{*}{3578}&\multirow{2}{*}{431,401}&\multirow{2}{*}{121}&\multirow{2}{*}{78}&\multirow{2}{*}{55}&\multirow{2}{*}{0.97}&\multirow{2}{*}{0.89}&\multirow{2}{*}{6}&\multirow{2}{*}{270}\\
w/ WCG&&&&&&&&&\\
\hline
\end{tabular}
\end{center}
\caption{Combined authnet characteristics with and without Windows Credential Guard (WCG). $|V|$: number of vertices; $|E|$: number of edges; $\overline{k_{\rm out}}$: average out-degree;  Trees: number of rooted trees (number of sets of reachable systems from $k_{\rm in} = 0$ vertices); Components: number of components; $|RGB|/|GCC|$: size of the residual giant bicomponent relative to the size of the giant connected component, obtained by removing the articulation point with the greatest reduction in size of the giant connected component; $\overline{\epsilon}$: average eccentricity; $d$: graph diameter; $\overline{|V_{{\rm desc}}|}$: average number of descendants of a vertex.}
\label{asan}
\end{table*}
\section{Finding Gatekeepers}
The combined authnet provides a near-complete and current map of all authenticated routes across the network.  It is therefore particularly useful for enumerating in real-time the possible pathways to a system or set of systems that are deemed especially critical assets.  These assets could be a service, like email or file shares, or the active sessions of highly-privileged accounts.  While cyber defenders might be monitoring these systems, by the time a compromise is discovered it might be too late.  It is therefore advantageous to anticipate the adversary by keeping a close eye on the known avenues of approach.  

In what follows, we refer to such critical asset as {\it targets}, under the presumption that an adversary would find them to be of high value.  To identify the key systems that facilitate access to targets, we first adopt a basic SI epidemic model as before.  The adversary can be assumed to be unguided, as in the spreader analysis, or more targeted: to model an adversary that knows where the targets are and seeks to minimize the number of hops, one can impose the constraint that the contagion follows the shortest paths from the initially infected host to the targets.   

\begin{figure}[ht]
\centering
\includegraphics[width=0.7\textwidth,clip]{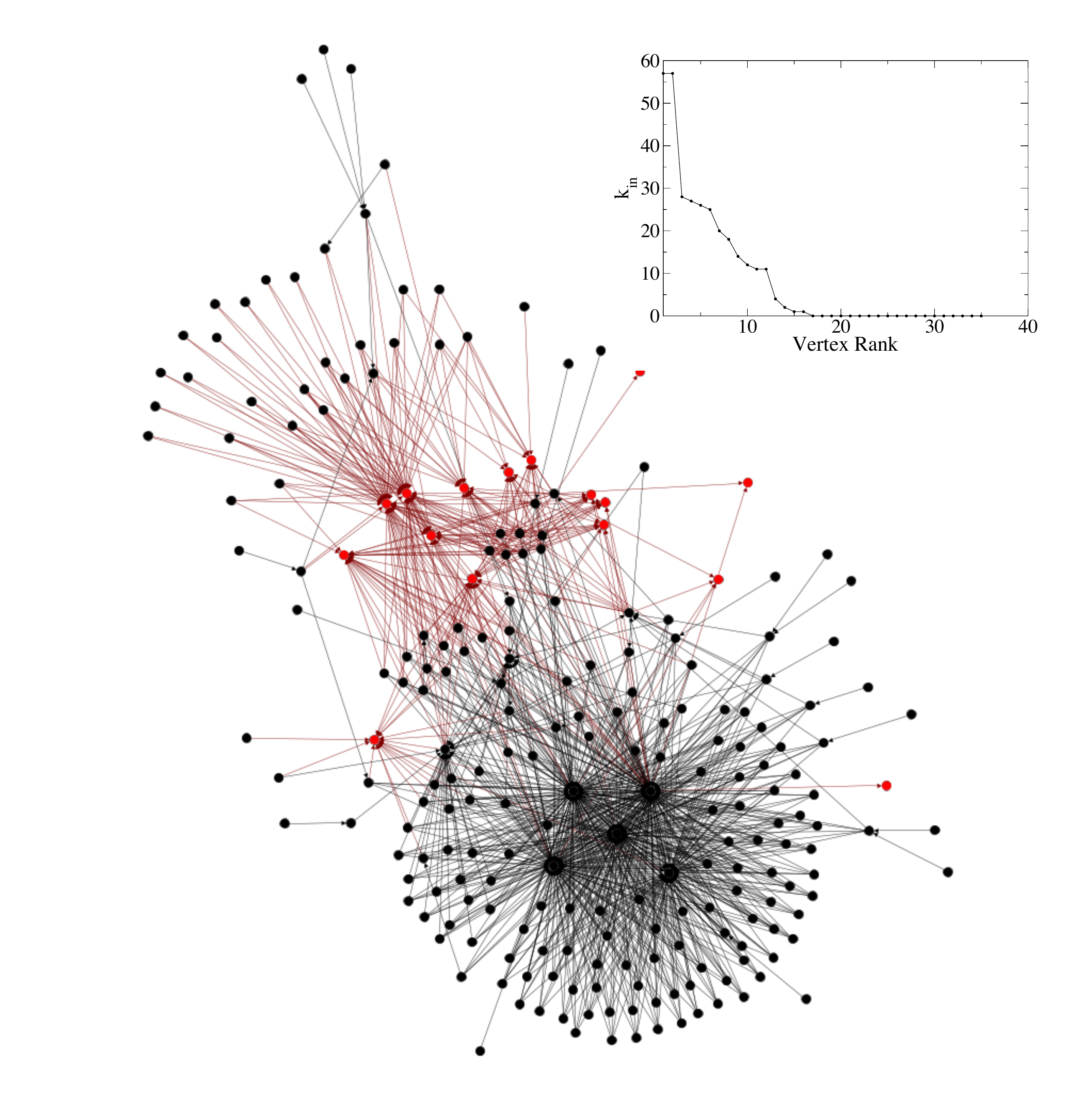}
\caption{Infection network including all systems that can reach 16 target systems with Domain Administrator sessions (red vertices).  Edges reveal the routes taken by the adversary following the shortest path from each system to the targets. Inset: plot of $k_{\rm in}$ for each of the 16 target systems, ranked from highest-to-lowest.}
\label{13}
\end{figure}

As a motivating example, we assume a Windows network and consider as targets all systems with active Domain Administrator sessions.  We assume a targeted adversary, and so we run the epidemic models on the combined authnet, restricted to the subgraph of all systems that can reach the targets along shortest paths.  As before, we run scenarios until all vertices in the associated authnet have been infected at least once, resulting in the infection network, Figure \ref{13}.  This network is easily derived directly from the combined authnet by retaining only shortest paths between each system and each target in the set.  

There are 35 Domain Administrator sessions, only 16 of which are reachable by some 230 other systems.  Some of these target sessions are directly reachable by a large number of systems, suggesting a simple exposure ordering based on $k_{\rm in}$, (inset, Figure \ref{13}).  For most of the targets, however, the exposure is not due to its neighbors, but its more distant ancestors.    Though a relatively small graph compared to the full authnet, there are likely too many systems here to monitor effectively\footnote{Defenders might monitor such systems by tracking logins, host processes, and outbound connections; by limiting the number of systems to control the volume of false positives, defenders can do a more thorough analysis of indicators and alerts.}.  We therefore want to find those systems through which the adversary is most likely to pass en route to the targets: these should comprise a relatively small and manageable set.  We refer to such systems as {\it gatekeepers}---in a strategy to combat the spread of a real disease, the individuals represented by these vertices would be good candidates for immunization.   

\begin{figure*}[htp]
\begin{tabular}{ccc}
\subfloat[Five gatekeepers (cyan vertices) found using Eq. (\ref{betT}).]{\includegraphics[width = 1.75in]{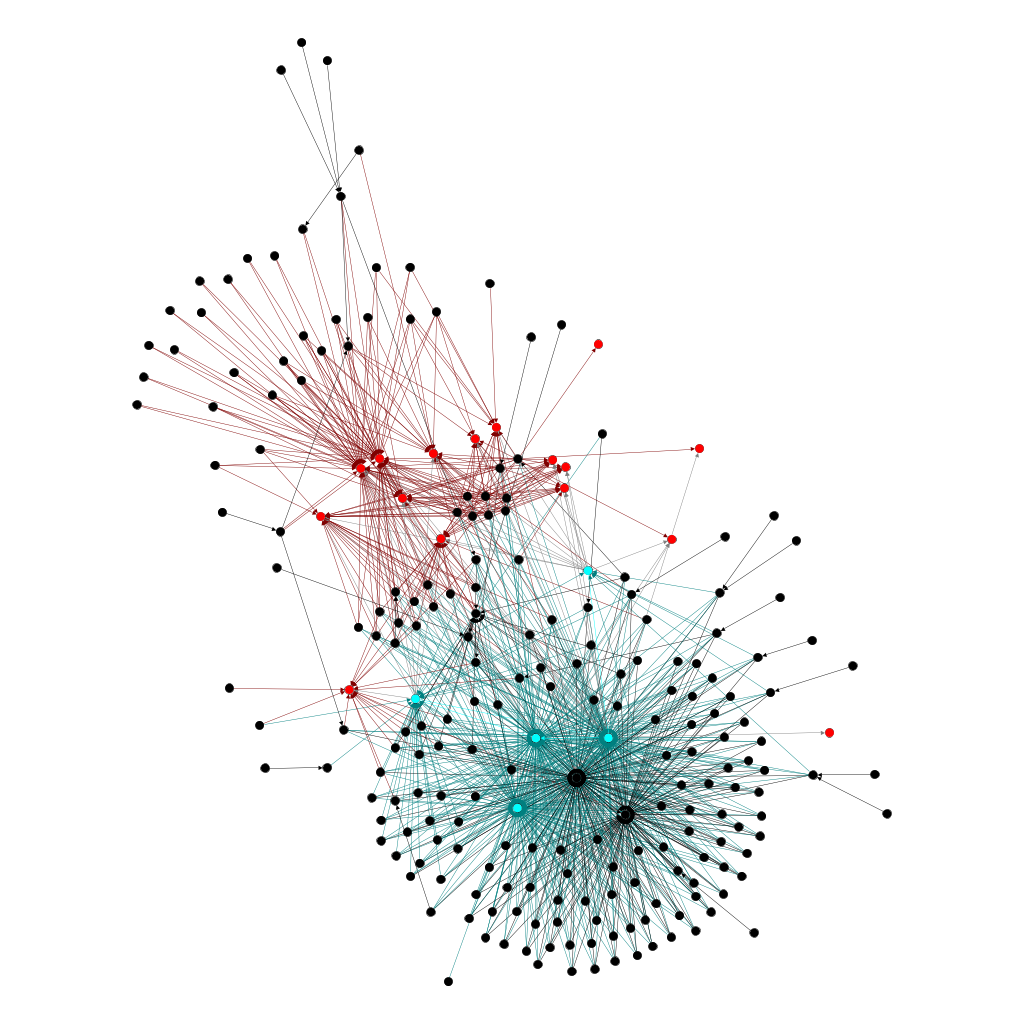}} &
\subfloat[Highlighting two vertices (blue): one a gatekeeper with modest $k_{\rm in}$ but reaching many targets, and one not a gatekeeper with large $k_{\rm in}$ but reaching only one gatekeeper.  See text.]{\includegraphics[width = 1.75in]{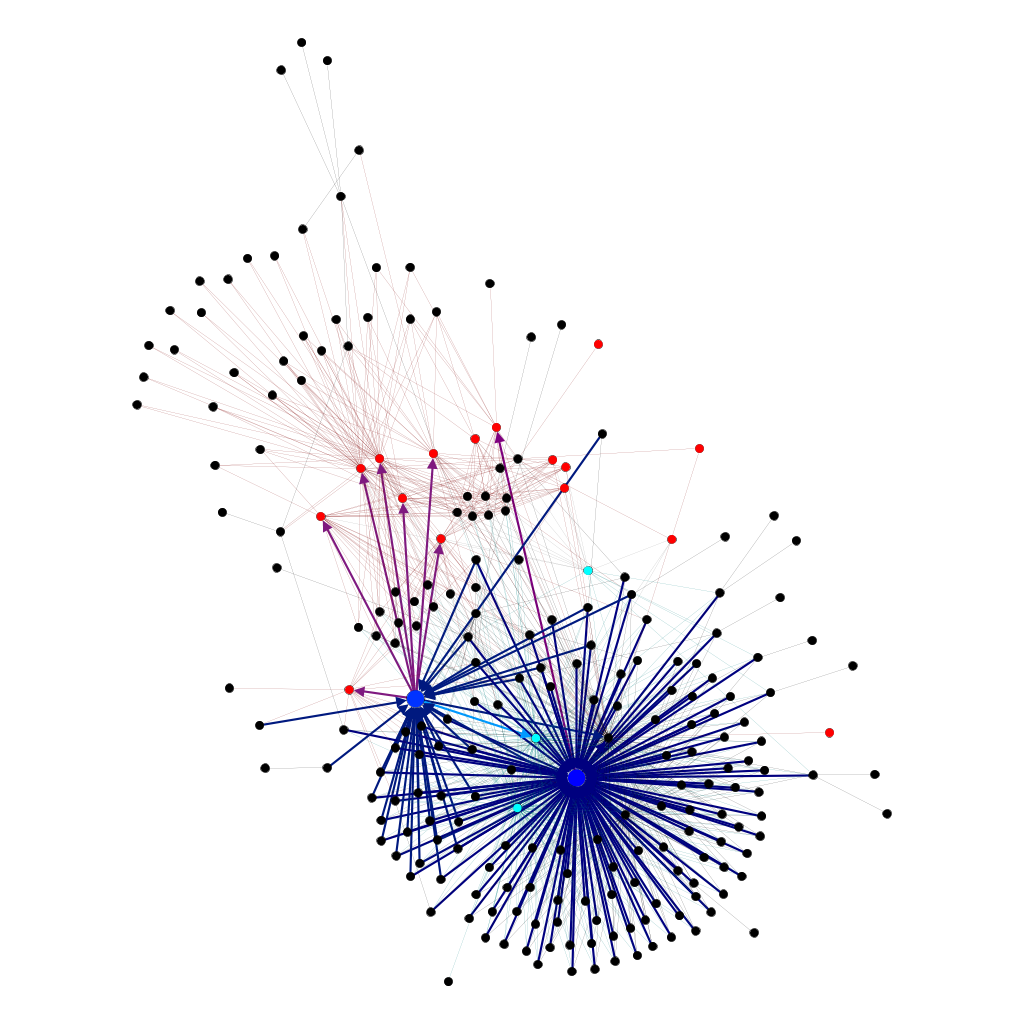}} & 
\subfloat[Seven gatekeepers (cyan vertices) found using Eq. (\ref{bet2}).]{\includegraphics[width = 1.75in]{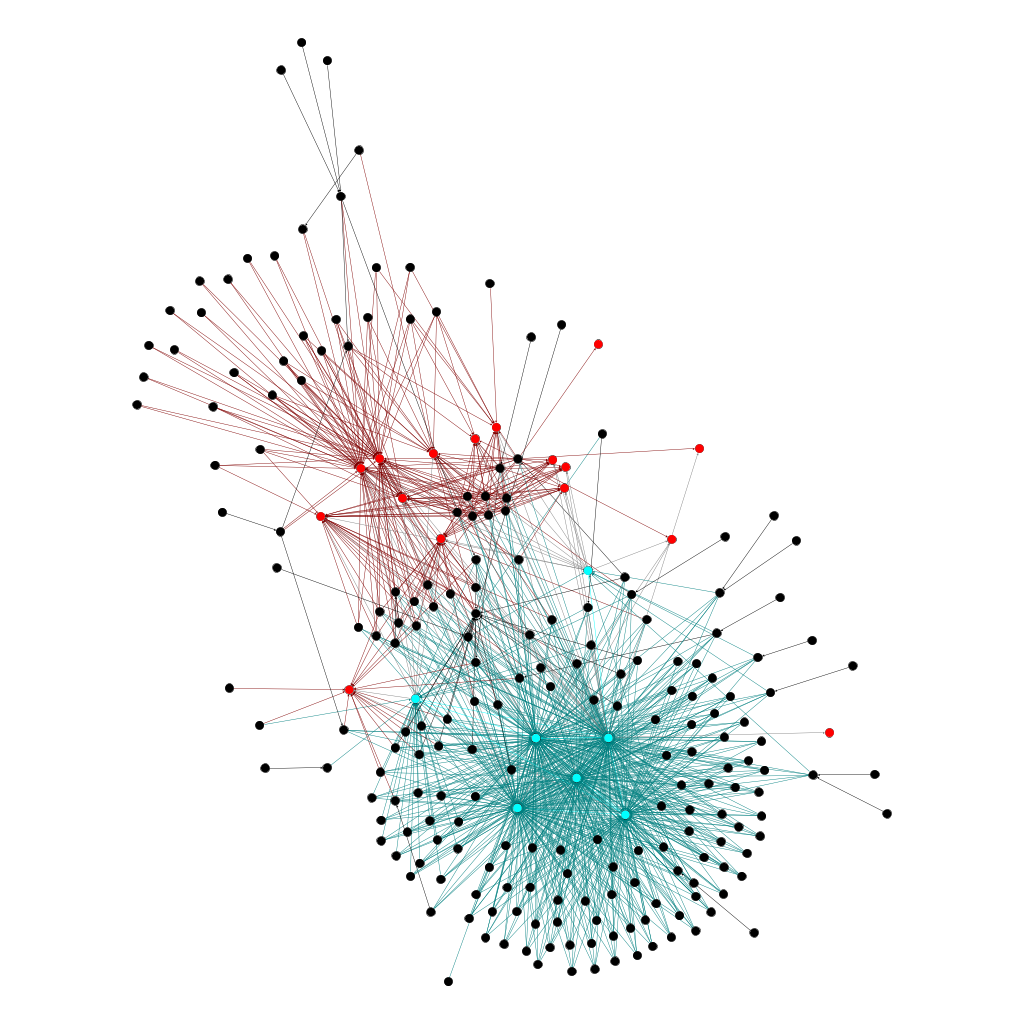}}\\
\end{tabular}
\caption{Gatekeepers of Domain Administrator sessions.}
\label{14}
\end{figure*}

The notion of gatekeeper vertices is by no means new \cite{Gould,Giuliani,Graf,Easley}.  Though definitions from social network theory vary, a common interpretation \cite{Easley,Acemoglu} is that a vertex $v$ is a gatekeeper if, given two other vertices $s$ and $t$, all paths between them pass through $v$.  Under this definition, being a gatekeeper is an all-or-nothing status.  In this analysis we relax this interpretation and instead associate it with shortest path betweenness centrality with respect to the target subset,
\begin{equation}
\label{betT}
b(v) = \sum_{s,t\in \mathcal{T}} \frac{n^v_{st}}{g_{st}} 
\end{equation}
where $n^v_{st}$ is the number of shortest paths between vertices $s$ and $t$ that pass through $v$, $g_{st}$ is the total number of shortest paths between $s$ and $t$, and where the endpoints, $t$, are restricted to the set of target vertices, $\mathcal{T}$.  If the infection network is not restricted to shortest paths as in this example, more general notions of betweenness can be applied, {\it e.g.} the communicability betweenness used earlier. Gatekeepers are identified as outliers in $b(v)$ with respect to other vertices in the target network: the distribution of $b(v)$ is not well fit by the standard parametric distributions and so we apply DBSCAN to find outliers as before.  The identified gatekeeper systems are colored cyan in Figure \ref{14} (a).  

These vertices ``intercept'' many paths destined for the target system, indicated by the blue edges in the figure.  But, there is something perhaps undesirable about this metric: notice that two of the vertices near the center of the large cluster are not found to be gatekeepers, though the other three similarly-situated vertices in the cluster are.  These five vertices have similarly large in degrees, meaning that they lie along the shortest paths of many of the same vertices destined to reach the targets.  The main difference is that the gatekeepers happen to connect to more targets than the two non-gatekeeper vertices.  Indeed, in Figure \ref{14} (b), we highlight a gatekeeper with $k_{\rm in } = 36$ but that connects to seven targets, and one of the two non-gatekeepers in the center of the cluster with a much larger $k_{\rm in} = 132$ but that only connects to a single target.  
However, in this example, where the targets are Domain Administrator sessions, it really does not matter how many sessions a given system can authenticate to: {\it only one is necessary!}  In this context, it seems we should care just as much about the highlighted vertex in Figure \ref{14} (b) that is not a gatekeeper as the one that is---perhaps even more so.  This undesirable feature stems from the fact that in Eq. (\ref{betT}) the betweenness receives contributions from all $t\in \mathcal{T}$.  An alternative definition that corrects for this shortcoming is the following, 
\begin{equation}
\label{bet2}
b(v)_{t_{\rm max}} = \max_{t\in \mathcal{T}}\sum_{s} \frac{n^v_{st}}{g_{st}} 
\end{equation}
where instead of summing over $t\in \mathcal{T}$ we select the $t$ for which $b(v)$ is a maximum.  With this new definition, we find seven outliers, including the two vertices left out under the definition Eq. (\ref{betT}), Figure \ref{14} (c).

\begin{figure*}[ht]
\begin{tabular}{cc}
\subfloat[Five gatekeepers (cyan) of single target (red) shown on SI infection network.]{\includegraphics[width = 2.7in]{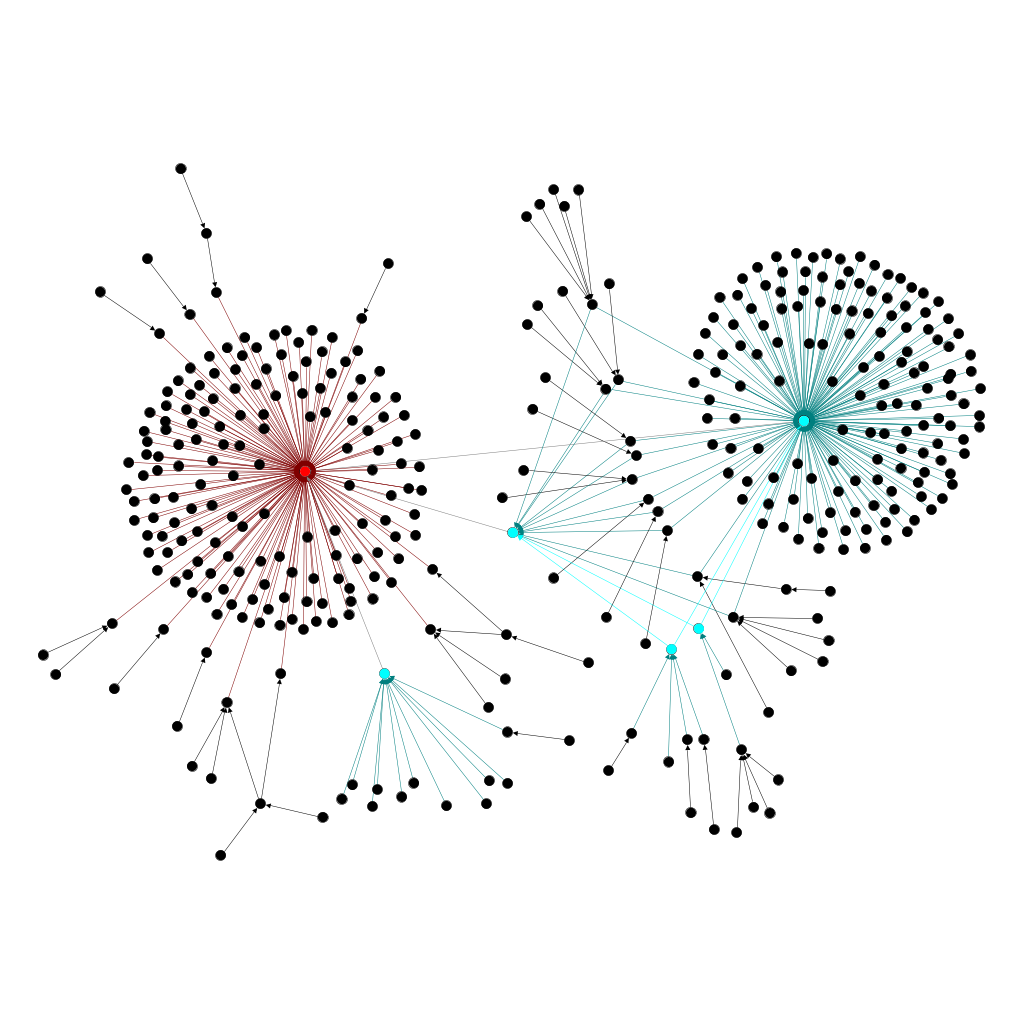}} &
\subfloat[Two gatekeepers (cyan) of single target (red) shown on SIR infection network.]{\includegraphics[width = 2.7in]{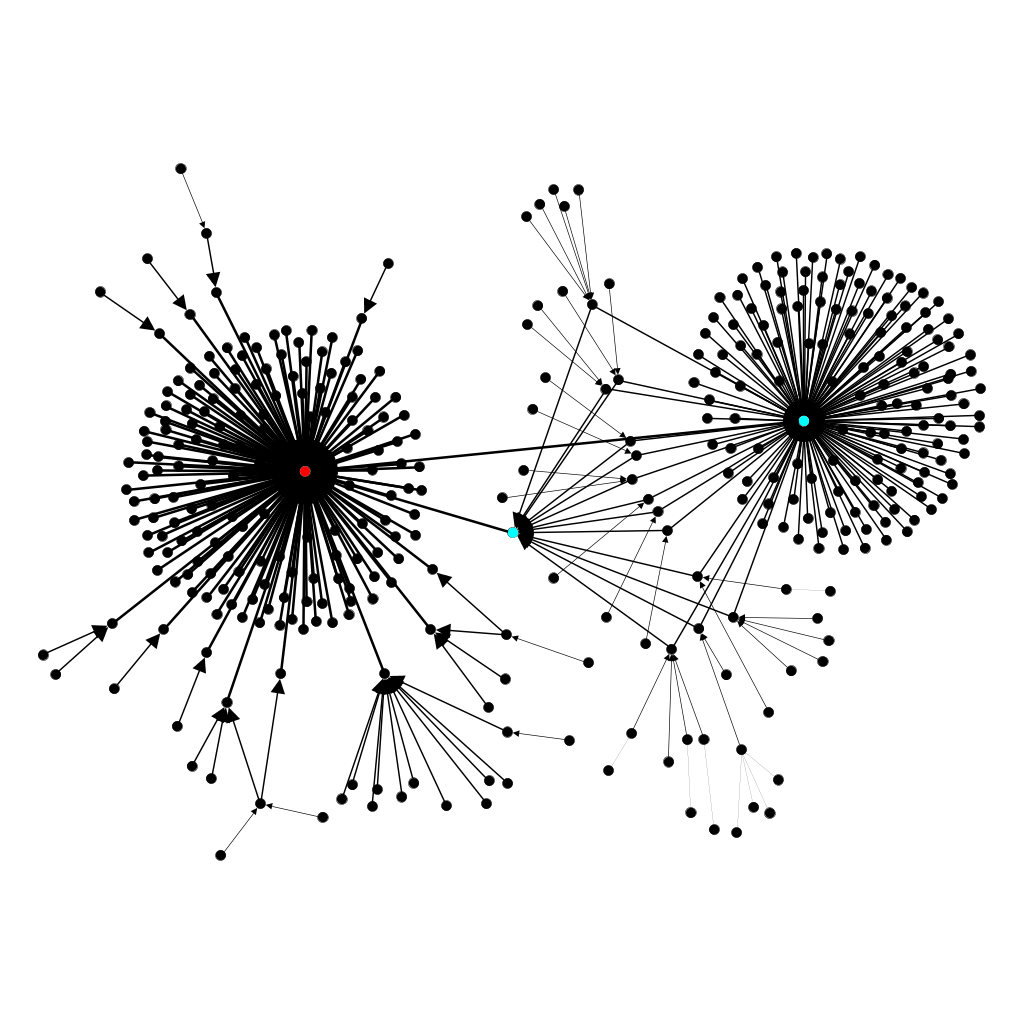}} \\
\end{tabular}
\caption{Gatekeepers of Exchange server.}
\label{15}
\end{figure*}
We next consider a more realistic model in which the adversary has a small probability of detection with each new access; this could be due to credential theft, reconnaissance activity, or anything else that happens to be noticed.  To account for the possibility of detection, we adopt an SIR epidemiological model in which recovery of a vertex corresponds to the discovery of the adversary on that system. We assume that a detected adversary is fully removed from the network, {\it i.e.} the infection scenario ends.  

To generate the infection network, we again consider an ensemble of epidemic scenarios covering all vertices in the target network.  As before, we run each infection until it can spread no further: if a vertex is ``cured'' before the contagion can spread to all reachable systems, the scenario is run again from all starting points until the epidemic completes from each one.  In this way, by the time the infection successfully reaches that targets from a distant starting system, there have been multiple scenarios completed from more close-by systems.  The resulting infection network resembles the associated authnet but with higher weight given to shorter paths to the targets.  

We demonstrate the SIR model on a different target network, one including all systems that can reach a single Exchange server.  We choose a 10\% detection rate\footnote{The detection rate need not be especially accurate, but can be tuned as an attenuation factor controlling how ``far out'' from the target set defenders should set their sights.} per access: this means that after two hops there is an 18\% chance of detection,  after three hops a 24\% chance, and so on.  The SIR infection network can be derived from the combined authnet as before, by restricting it to include only the shortest paths between the targets and ancestor systems.  We assign a weight to edges as follows: for an infection starting at vertex $s$ that is $n$ hops from the target, $t$, each edge $(x_i,x_{i+1})$ along this path receives a weight $w_{i,i+1} = p^{n-1}$.  In cases where the same edge belongs to more than one path, it assumes the weight of the highest-weight path.   To find the gatekeepers, we employ a weighted betweenness measure
\begin{equation}
%b(v)_{t_{\rm max},w} = \max_{t\in \mathcal{T}}\sum_{s} \left(\frac{\sum_i^{\ell^{v}_{st}-1}w^v_{i,i+1}}{\sum_i^{\ell_{st}-1}w_{i,i+1}}\right) 
b(v)_{t_{\rm max},w} = \max_{t\in \mathcal{T}}\sum_{s} \frac{w_{st}^v}{w_{st}}
\label{betw}
\end{equation}
where $w_{st}^v$ is the sum of the edge weights along all shortest paths from $s$ to $t$ that pass through $v$, and $w_{st}$ is the sum of the edge weights along all shortest paths between $s$ and $t$. For a given path of length $d(s,t)$,
\begin{equation}
w_{st} = \sum_i^{d(s,t)-1} w_{i,i+1}
\end{equation}
where $w_{i,i+1}$ is the weight of the edge $E = (x_i,x_{i+1})$, $s = x_1$, and $t = x_{d(s,t)}$.  

The gatekeepers for this network are shown in Figure \ref{15} (b); for comparison, in Figure \ref{15} (a) we include the gatekeepers for the corresponding network without the possibility of detection (SI model)\footnote{Note that two of the gatekeepers are fully ``blocked'' by two others in Figure \ref{15} (a) and so would ordinarily not be considered; we include them here to illustrate how the possibility of detection reduces the centrality and significance of these more distant vertices.}.  Fewer gatekeepers are needed at the same outlier threshold for the SIR infection scenario, since long paths are attenuated.  The SIR model is most practical in resource-scarce environments, where defenders can only afford to focus on a small subset of the most critical conduit systems---the SIR model directs this focus to key systems one or two hops out from the targets.  

\section{Conclusions}
An analytical approach has been presented that is useful for identifying systems and accounts at risk of compromise by an adversary moving laterally through a computer network.  The adversary was modeled as an epidemic spreading through the network of authentication relationships via credential chaining.  These relationships take the form of a directed graph, called the authentication network, derived from Administrator access and active session data, which together provide a comprehensive and timely map of credential exposure across the network.   The infection network that results from running epidemic scenarios on the authnet was analyzed using graph centrality measures and other characteristics to identify critically exposed systems.  Finding such systems is essentially an outlier detection problem: density-based clustering is used to find exceptional vertices in the multi-dimensional centrality space.  

We applied this approach to study three sources of exposure.  First, we employed an SI epidemiological model to identify systems, which, either independently or collectively, provide wide and far-reaching access to many other systems across the network: these {\it spreaders} are efficient facilitators of lateral movement and are thus of value to the adversary. Two different countermeasures were investigated: we employed community detection to determine how access controls might be tightened on individual accounts to reduce their spreadability, and next the global policy of disabling remote access for local accounts was analyzed for its effect on spreadability. We found that, on networks where feasible, this global restriction has a significant effect on spreadability, reducing the number of descendants  and eccentricity of top spreaders by as much as 97\% and 80\%, respectively.  More generally, the proposed framework allows analysts to explore the effectiveness of different countermeasures against different threat models without the need for test deployments.

Next, we identified connected systems with active sessions by Administrative accounts in different security tiers, indicating the potential for privilege escalation.  These {\it escalators} are characterized by the number of new accesses that the privilege escalation grants, which can be used to ascertain significance depending on the threat model: escalators with many connections, only a few of which result in a large increase in access, are more relevant to targeted adversaries that know where they are going; escalators with more modest access expansion spread more evenly over its connections are of more concern against a random adversary.  This analysis can be used to identify individual Administrator accounts that frequently cross security tiers when accessing systems remotely, suggesting either an access control problem or signaling a need to schedule remote logins so that cross-tier sessions are not mutually accessible.  Additionally, this methodology can be used to explore how well countermeasures that limit credential residue on systems reduce the significance of escalators.  We found that under the assumption that Windows Remote Credential Guard, which reduces credential residue for remote interactive logins, is installed on all systems and found that active sessions with privileged credential residue were reduced by 25\%, and that fewer escalators could reach the highest-privileged accounts.  

Lastly, we applied this approach to the more reactive problem of real-time cyber defense.  Given a set of key defendable assets, we identified {\it gatekeepers}---conduit systems through which an adversary must pass en route to these assets.  We employed an SIR epidemic model, with recovery corresponding to detection, to more realistically characterize the adversary in the defensive setting.  Systems with high shortest path betweenness centrality with respect to the target systems were identified via outlier analysis as gatekeepers---these systems lie at the frontier of the cyber attack and so are good candidates for the placement additional ``early warning'' defensive resources.  

This methodology has been demonstrated on a Windows network, but the approach is general once the authentication network is built.  For Windows Domains, the data necessary to build the authentication network can be acquired via common Domain queries and from Windows Event Logs describing logon, logoff, and reboot events.  Analyses relevant to Active Directory configuration and Domain-wide mitigations need only to be conducted periodically, where the effectiveness of different mitigations can be explored by studying how well contagion spreads on the corresponding authnets.  The results of this kind of study should provide analysts with the ability to make cost/benefit decisions regarding large-scale mitigations or configuration changes.  Meanwhile, analyses of gatekeeper systems are more relevant to the daily tempo of defensive cyber operations: as active sessions come and go, there is a shifting landscape of authenticated routes to targets.  Logs can be queried daily or even hourly in order to maintain an up-to-date authnet for with current gatekeeper candidates.  

Because it is so quiet and tends to look so normal, authenticated lateral movement remains a daunting defensive challenge.  On large networks with many privileged accounts, there is a lot of noise and little signal, and generally only the known critical systems are observed closely by defenders.  This work seeks to improve prevention by helping analysts tighten access configurations and test protective measures, so that that the adversary has a harder time moving around the network and escalating privileges.  This work also seeks to improve detection by cueing  network defenders on where to look for adversarial accesses in order to better anticipate movement towards critical systems.  While running the correlations and crafting alerts to detect stealthy lateral movement and credential theft remains the utmost challenge, the input from this analysis can help defenders scope these activities to the most exposed systems on the network. 

\section*{Acknowledgement}
The author would like to thank colleague Max Kresch for providing valuable comments on a draft version of this article.

\bibliographystyle{tfnlm}
%\bibliography{interactnlmsample}
\bibliography{draft}

\appendix
\section{Authentication and Credentials}
In this Appendix, we provide some details on the various authentication technologies employed on Windows systems and Domains.  
\subsection{Users and Groups}
A {\it user account} is the digital identity of a user in relation to a computer or network.  In addition to an identifier and an authenticator, like a password, the user account declares the authorizations of the user, including access controls, permissions, and so on. A user account can be either local to a computer, for which authentication takes place against the computer's own security database, or it can belong to the Domain, in which case authentication takes place against the Domain's centralized database.  The centralized database is maintained by the Domain Controller, which provides account and directory services across the Domain via the Microsoft Active Directory service.  

A {\it group} is a collection of users sharing common privileges and access controls. Like user accounts, groups can be local to a system or can apply at the Domain level.  Groups are a useful way of controlling the access and privilege of a large number of similar accounts, and for controlling user privileges that apply to a large number of systems in the Domain. 

One important group, the Local Administrators, gives members full administrative control of the system.  Every system generally has a ``built-in'' Administrator account in this group; often ordinary user accounts, like that of the system owner, are added to this group as well.  When a collection of users needs Local Administrator access to many systems, common practice is to gather these users together into a Domain group, and then place this group into each system's Local Administrator group.  The Domain Administrators group is one such group, the members of which generally have Local Administrator privileges on all systems within the Domain, in addition to key components like Domain Controllers.  Lower-tier Administrators, like desktop or database Administrators, generally belong to groups with more restrictive access or with fewer privileges on authorized systems. 
\subsection{Logon Types}
Information systems can be accessed either locally or remotely over a network.  In Microsoft parlance, an {\it interactive} logon occurs when a user enters a credential either at the local or remote console of a computer; these are called Microsoft Type 2 and Type 10 logons, respectively.  Interactive logons give the user the ability to provide input to and receive output from the system, and to access additional resources from the remote system.  In contrast, a {\it network} logon (Type 3) occurs after an interactive logon to the system initiating the network logon, and provides the user access to certain shared resources, like files or printers, over the network.  This authentication happens in the background, making use of the user credential that was provided as part of the initial interactive logon.  

In order to support additional access and resource requests after interactive logon, host systems must have access to user credentials. Microsoft Single Sign-on conveniently automates all subsequent authentications so that users aren't prompted for a password each time a new resource is accessed.  For this to work, the user's credentials must be made available to the range of security service providers (SSP) that establish these accesses, and so are placed in the running memory allocated to the Local Security Authority Server Service (LSASS) process that controls the SSPs.  For remote interactive logons, this means that credentials might be sent over the network and placed in the remote system's running memory.   
\subsection{Impersonation Levels}
When a user authenticates to a remote server, they might require access to restricted services or other remote hosts. This requires that the remote server negotiate access on the client's behalf.  The extent to which the remote server can do this is formalized in terms of four {\it impersonation levels}: {\it anonymous}, where the remote server has no ability to act on the client's behalf (it has no information about the identity of the client user); {\it identification}, where the remote server can obtain the client's identity for the purpose of doing access control checks; {\it impersonation}, where the remote server can access local resources as the client; and {\it delegation}, where the remote server can access remote resources as the client.  Delegation might involve the remote server authenticating as the client to other remote systems, and can be constrained to apply to only specific services.   
\subsection{Authentication Protocols} 
Windows systems use two primary authentication protocols: NTLMv2 and Kerberos. NTLMv2 was introduced as the default authentication package with Windows NT4.0 in 1996: it is a cryptographic challenge/response that makes use of the client user's NT hash\footnote{The NT hash is the unsalted MD4 hash of the user's UTF-16-encoded password.} as the key.  
The NT hash can be used for both local and Domain logins: in the event of local login, the NT hash is verified against the system's password database, the Security Account Manager (SAM); for Domain, NTLMv2 is used over the network to authenticate the user against the Domain Controller's security database.  After authentication, the NT hash is placed into LSASS memory on the local system for subsequent operations that require impersonation.  NTLMv2 can also be used to support network and remote interactive logons, like Remote Desktop Protocol (RDP); after remote interactive authentication, the NT hash is placed in the remote system's LSASS process memory.  On Domain-joined systems, the NT hashes of non-local accounts that access the system via NTLMv2 are cached in the registry so that these accounts can authenticate again in the future if the Domain Controller is not available. 

Since Windows 2000, the default package for Domain authentication has been {\it Kerberos}, which is a centralized service that employs symmetric encryption to authenticate users and control access to requested services (see Appendix A).  When the user contacts the Kerberos Authentication Server, they receive encrypted data called the {\it ticket-granting ticket} (TGT) and a session key; the user then provides the TGT to another Kerberos component called the Ticket-Granting Server (TGS) along with a requested service, like a file share or email.  The TGS then provides the user with a {\it service ticket}, which is an encrypted session key for use with the requested service.  These tickets are all stored in LSASS for a prescribed duration (default is seven days for TGT, renewed every ten hours).  To summarize, there are three Kerberos artifacts related to authentication available on the client: the TGT, the TGS session key, and any service tickets.  For network and remote interactive logins, ordinarily only pertinent service tickets are stored in memory on the remote host; however, if the user requires access to additional services from the remote server ({\it e.g.} if they wish to access a back-end database), then the TGT and TGS session key are stored on the remote server for delegation.

\begin{table*}[htp]
\begin{center}
\footnotesize
\begin{tabular}{|c|c|c|c|c|c|c|}
\hline
Logon&Authentication&Imp&\multirow{2}{*}{Local}&\multirow{2}{*}{Remote}&Local w/&Remote w/\\
Type&Package&Level&&&WCG&WCG\\
\hline
\multirow{4}{*}{2}& \multirow{2}{*}{NTLMv2}&Id&Hash&\cellcolor{gray}&-&\cellcolor{gray}\\
%\cline{3-7}
&&Imp&Hash&\cellcolor{gray}&-&\cellcolor{gray}\\
\cline{2-7}
&\multirow{2}{*}{Kerberos}&Id&TGT,ST,TGS SK&\cellcolor{gray}&-&\cellcolor{gray}\\
%\cline{3-7}
&&Imp&TGT,ST,TGS SK&\cellcolor{gray}&-&\cellcolor{gray}\\
\hline
\multirow{5}{*}{3}& \multirow{2}{*}{NTLMv2}&Id&Hash&-&Hash&-\\
%\cline{3-7}
&&Imp&Hash&-&Hash&-\\
\cline{2-7}
& \multirow{3}{*}{Kerberos}&Id&TGT,ST,TGS SK&-&TGT,ST,TGS SK&-\\
%\cline{3-7}
&&Imp&TGT,ST,TGS SK&ST&TGT,ST,TGS SK&ST\\
%\cline{3-7}
&&Del&TGT,ST,TGS SK&TGT,ST,TGS SK&TGT,ST,TGS SK&ST\\
\hline
\multirow{5}{*}{10}& \multirow{2}{*}{NTLMv2}&Id&Hash&-&Hash&-\\
%\cline{3-7}
&&Imp&Hash&Hash&-&-\\
\cline{2-7}
&\multirow{3}{*}{Kerberos}&Id&TGT,ST,TGS SK&-&-&-\\
%\cline{3-7}
&&Imp&TGT,ST,TGS SK&ST&-&ST\\
%\cline{3-7}
&&Del&TGT,ST,TGS SK&TGT,ST,TGS SK&-&ST\\
\hline
\end{tabular}
\end{center}
\caption{Credential residue by Logon Type, Authentication Package, and Impersonation Level, on hosts with and without Windows Credential Guard (WCG).  TGT = ticket-granting ticket, ST = service ticket, TGS SK = ticket granting server session key, Id = Identification, Imp = Impersonation, Del = Delegation.}
\label{results}
\end{table*}
\section{Credential Theft and Lateral Movement}
Successful interactive logon to a Windows system generically involves three important steps vis-\`{a}-vis credentials: the user provides a credential that is verified by a security authority, the user is granted access to the information system and credentials are placed into running memory for Single Sign-on, and, for Domain logons,  the user's credential is cached by the local security authority for future authentications in the event that the Domain security authority is unavailable. We highlight these three stages because each involves the location of a user's credentials: primary security database, running memory, and cache---these are the spots the adversary will search for credentials.  

Adversarial tradecraft for the theft of user credentials is wide and deep, with different tools and techniques depending on the types and storage locations of credentials.  The following is a brief overview of common methods and tools.  
\subsection{Local SAM and Registry}
When a Windows system is running, the SAM and related registry files are locked and cannot be read or accessed by non-SYSTEM entities.  If an adversary is able to escalate privileges to SYSTEM, however, the native tool \texttt{reg.exe} can be used to easily interact with the registry. 
Prior to Windows 10, sensitive data in the SAM is encrypted with a key (called the {\it Syskey}) derived from data found in the \texttt{SYSTEM} registry hive: tools like \texttt{bkhive} can extract the Syskey and then tools like \texttt{samdump2} use it to decrypt the SAM and dump the NT hashes\footnote{With Windows 10, the SAM encryption scheme has changed and no longer uses Syskey, rendering these approaches ineffective.}.  Metasploit Meterpreter's \texttt{hashdump} post-exploitation module combines these two actions.  
In a similar vein, cached Domain hashes can also be decrypted and dumped from the registry: tools like \texttt{cachedump.py} \cite{cachedump} can recover them but they are heavily obfuscated with PBKDF2 and must be cracked to obtain the NT hash.

If escalating to SYSTEM-level privileges is a problem, it is possible to circumvent the filesystem locks on the registry entirely by reading data straight from the disk volume with only Local Administrator rights.  The tool \texttt{pwdump7} \cite{pwdump7} uses its own filesystem drivers in place of the standard Win32 API file read functions that are subject to filesystem locks; it can decrypt and dump SAM hashes in one execution.  Altneratively, the tool \texttt{ninjacopy} \cite{ninjacopy} is a PowerShell utility with its own NTFS parser so that it too can dispense with Win32 API calls. The rub is that \texttt{pwdump7} must be loaded onto the target system, potentially triggering antivirus, and, while \texttt{ninjacopy.ps1} can be launched remotely, the script must be injected into running memory as a dynamic-link library (DLL), which  might trigger alarms. 
\subsection{Memory}
Perhaps the most common route to credentials is through running memory.  Microsoft Single Sign-on ensures that a user's credentials are available in LSASS memory during active sessions if they are needed for impersonation or delegation.  In additional to hashes and Kerberos artifacts, plaintext passwords are stored for use by other authentication pacakges, like Windows digest and terminal services. The tool \texttt{mimikatz} \cite{mimikatz} has been wildly successful at retrieving each of these credentials from LSASS.  It operates either with or without DLL injection into LSASS, and can even operate on a static LSASS memory dump (obtained, for example, with the Sysinternals tool ProcDump \cite{procdump}).  Since Windows 8.1, credentials are no longer stored in memory by default; however, this behavior can be changed by altering the registry.  This requires SYSTEM-level privileges, but so too does injecting into or otherwise interacting with LSASS memory, and so this safeguard does not pose any additional challenges to an attacker.

If hashes alone will suffice, Metasploit's \texttt{hashdump} run as a Meterpreter command and older executables in the \texttt{pwdump*} family and its variants retrieve user hashes from memory.

\subsection{Remote Access}
Lateral movement involves the remote access of one system from another. 
%generally as an interactive session so that commands can be run on and output received from the remote system.  
Methods of remote access vary, and which is chosen depends on host configuration, availability of tools, desired level of stealth, and possibly the type of credential available for use.  Prior to establishing remote access with a compromised account, the adversary will generally use the stolen credential to create a user access token for the compromised account on the source system.  This can be done with a stolen hash (called ``pass-the-hash'') with tools like Windows Credential Editor or \texttt{mimikatz}, or a Kerberos TGT (called ``pass-the-ticket'') with \texttt{mimikatz}.  The NT hash can also be used to request TGTs from the TGS (called ``overpass-the-hash''). Once the user access token is created, the adversary can simply use native Windows utilities to authenticate to and remotely access the target system.  Useful native tools for remote access include PsExec, Windows Management Instrumentation Command-Line Utility (WMIC), Windows Remote Manager (WinRM), and Distributed Component Object Model (DCOM) communication. These technologies allow for the execution of remote processes, and so are quite general in their capabilities.  For example, if an interactive session is desired, they can be used to spawn a \texttt{cmd.exe} shell or PowerShell console.  

Interactive login can also be achieved with RDP, telnet, or SSH if available.
\section{Kerberos Protocol}
\begin{figure*}
\centering
\includegraphics[]{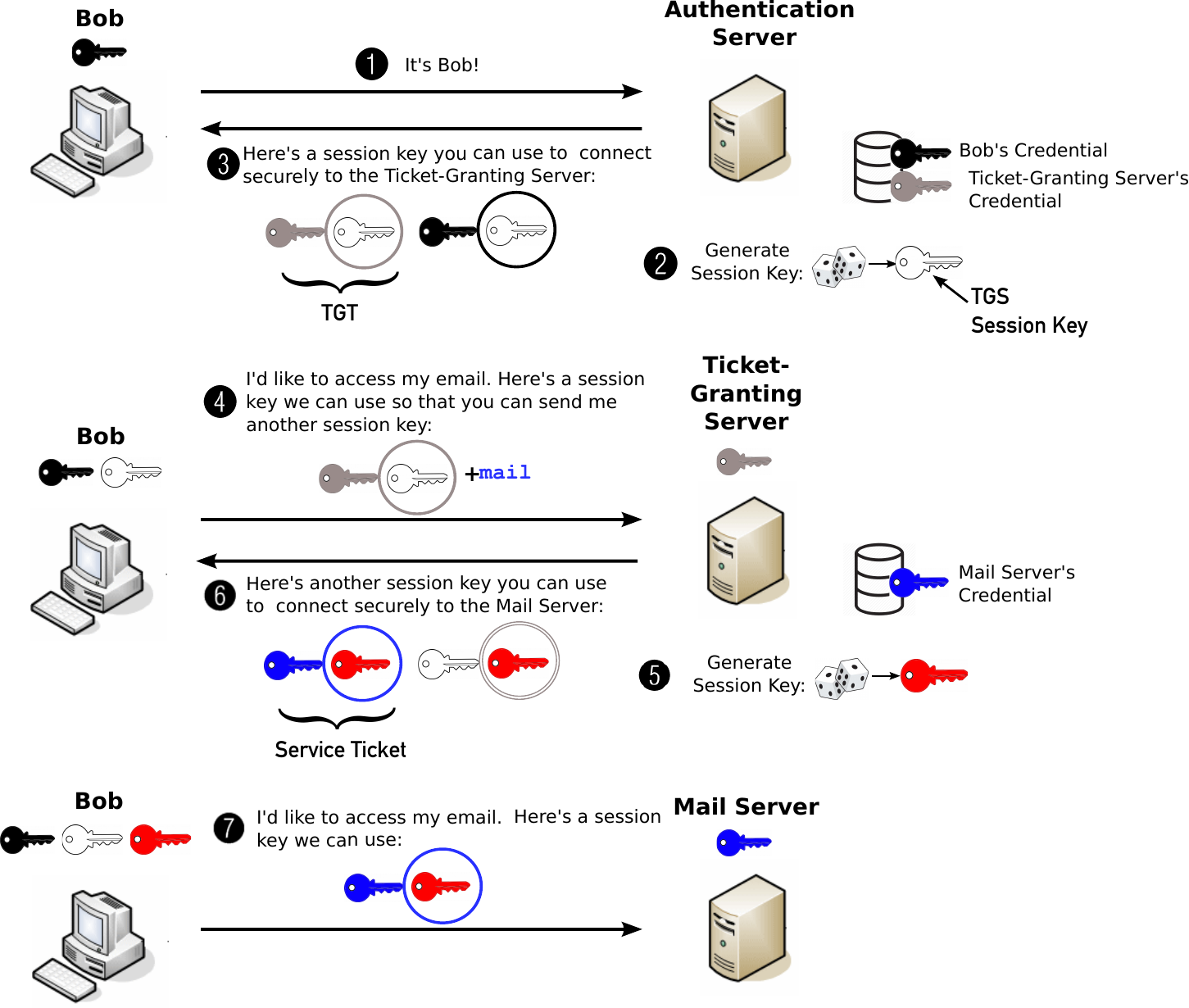}
\caption{Brief schematic of the various Kerberos key exchanges required to gain access to a server. See text for details.}
\label{16}
\end{figure*}

Figure \ref{16} is a simplified diagram of the Kerberos protocol, emphasizing the various key exchanges relevant to credential theft.  1) Bob requests a ticket-granting-ticket (TGT) from the Authentication Server. 2) The server generates a random session key (the Ticket-Granting Server (TGS) session key) and encrypts one copy with the TGS's key (this is the TGT) and one copy with Bob's key, and 3) sends them both to Bob.  Bob can decrypt the session key with his own secret key and keeps it handy. 4) Bob requests a service (mail) from the TGS by sending the TGT to the TGS.  The TGS can decrypt the TGT and retrieve the TGS session key. 5) The TGS generates a random session key and encrypts one copy with the TGS session key and one copy with the mail server's key (this is the service ticket).  6) Both of these tickets are sent to Bob, who uses the TGS session key to decrypt the session key.  7) Bob sends the service ticket to the mail server, who decrypts it and retrieves the session key.  Both Bob and the mail server can now use the session key for secure communications.  Authentication is indirectly achieved since the encrypted session keys can only be decrypted and used by entities in possession of the intended recipient's credential.
\end{document}